\documentclass[a4paper,fleqn,usenatbib,10pt,twocolumn]{article}

\usepackage{newtxtext,newtxmath}
\usepackage[T1]{fontenc}
\usepackage{ae,aecompl}

\usepackage{graphicx}	
\usepackage{amsmath}	
\usepackage{amssymb}	
 \usepackage{array}      
\usepackage{graphicx}				
\usepackage{amssymb}
\usepackage{pdfpages}
\usepackage{wrapfig,lipsum,booktabs}
\usepackage{subcaption}
\usepackage{float}
\usepackage{breakcites}
\usepackage{soul}
\usepackage{hyperref}

\usepackage[margin=0.7in]{geometry}

\title{Gaian bottlenecks and planetary habitability maintained by evolving model biospheres: The ExoGaia model}

\author{
Arwen E. Nicholson,$^{1}$\thanks{E-mail: arwen.e.nicholson@gmail.com}
David M. Wilkinson,$^{2}$
Hywel T. P. Williams,$^{1}$
\\
and Timothy M. Lenton$^{1}$
\\
\\
$^{1}$Earth System Science, University of Exeter, UK\\
$^{2}$School of Life Sciences, University of Lincoln, UK
}

\date{Accepted for publication by MNRAS 12th March 2018}

\begin{document}

\label{firstpage}
\twocolumn[
  \begin{@twocolumnfalse}
    \maketitle
\begin{abstract}
The search for habitable exoplanets inspires the question - how do habitable planets form? Planet habitability models traditionally focus on abiotic processes and neglect a biotic response to changing conditions on an inhabited planet. The Gaia hypothesis postulates that life influences the Earth's feedback mechanisms to form a self-regulating system, and hence that life can maintain habitable conditions on its host planet. If life has a strong influence, it will have a role in determining a planet's habitability over time. We present the ExoGaia model - a model of simple `planets' host to evolving microbial biospheres. Microbes interact with their host planet via consumption and excretion of atmospheric chemicals. Model planets orbit a `star' which provides incoming radiation, and atmospheric chemicals have either an albedo, or a heat-trapping property. Planetary temperatures can therefore be altered by microbes via their metabolisms. We seed multiple model planets with life while their atmospheres are still forming and find that the microbial biospheres are, under suitable conditions, generally able to prevent the host planets from reaching inhospitable temperatures, as would happen on a lifeless planet. We find that the underlying geochemistry plays a strong role in determining long-term habitability prospects of a planet. We find five distinct classes of model planets, including clear examples of `Gaian bottlenecks' - a phenomenon whereby life either rapidly goes extinct leaving an inhospitable planet, or survives indefinitely maintaining planetary habitability. These results suggest that life might play a crucial role in determining the long-term habitability of planets.\newline \newline
$^{*}$ arwen.e.nicholson@gmail.com \newline
\end{abstract}
\end{@twocolumnfalse}
]

\section{Introduction}

Most models of habitable planets and the boundaries of the habitable zone focus on the physical processes happening on planets to determine the limits of habitability (for example \cite{Cockell2007} and \cite{Kopparapu2013}). These models neglect a biotic response to changing conditions on an inhabited planet. The Gaia hypothesis postulates that life influences the Earth's feedback mechanisms to form a self regulating system \cite{Lovelock1974, Lenton1998, Lovelock2000}. We see the signature of life on our planet in the chemical composition of our atmosphere, oceans, and soil. If life has a large effect on its host planet, this has implications for habitable exoplanet research. One area where the Gaia hypothesis has relevance for exoplanet research is around the establishment of life on a previously uninhabited planet. The idea of Gaian bottlenecks \cite{Chopra2016} suggests that early in a planet's history, assuming initially habitable conditions, life must quickly establish self regulating feedback loops in order to maintain habitable conditions. If it fails, life goes extinct and leaves the planet in a lifeless state. Gaian bottlenecks could be linked to recent models of bifurcations in early planet formation \cite{Lenardic2016}.  

\cite{Lenardic2016} suggest that the end state of a planet is not entirely deterministic. Plate tectonics, key to climate regulation, are affected by the temperature of the planet. To demonstrate this, Lenardic et. al. simulate a planet with plate tectonics, heat the planet until the tectonics disintegrate and then cool the planet back to its original temperature. After cooling, the plate tectonics are not reformed, suggesting that there are two stable states for the same temperature. Venus and Earth are traditionally thought to be different classes of planet - although similar in size, mass and chemical composition, Venus's closer orbit to the Sun is thought to have doomed it to its current  state. However, evidence suggests that Venus once hosted large bodies of water \cite{Donahue1982, Jones2003}, which since boiled away in a runaway greenhouse effect (\cite{Kasting1988}), leading to present day surface temperatures of over 400$^{o}$C. Lenardic et. al. suggest that Earth and Venus could represent two different stable states for the same system. They predict that small fluctuations early in a planet's history could determine the end state of that planet, with life possibly providing such a perturbation. Earth is very different from what it would be if it were uninhabited. Our atmosphere would be dominated by $CO_{2}$ as is the case on Venus and Mars and this would affect the Earth's surface temperature. If Venus once had life back when it had water, could we be on the lucky side of a Gaian bottleneck? While most models place Venus outside the habitable zone as Venus receives almost twice the amount of solar radiation as Earth, a few allow the potential for a habitable Venus today (e.g. \cite{Zsom2013} and \cite{Yang2014}). Recent models (\cite{Yang2014, Way:2016aa}) demonstrate the important role planetary rotation and topography play in understanding a planet's climatic history, and suggest that rocky planets that retain significant water after formation can experience habitable conditions well within the traditionally defined inner edge of the habitable zone (e.g \cite{Kopparapu2013}).

Inspired by these important questions, we present a new abstract model of environmental regulation performed by evolving biospheres - the ExoGaia model. We model simple `planets' with atmospheres whose chemical composition influences planetary temperatures. Model microbes consume and excrete atmospheric chemicals, via temperature dependant metabolisms. Thus microbes can impact planetary temperatures by altering the chemical composition of their host planet's atmosphere. We investigate whether a simple biosphere can regulate its host planet's temperature within habitable bounds. We focus on a biotic response to planetary conditions, in contrast to most habitability models. We do not attempt to model the complexities of real planets, allowing us to isolate purely biotic phenomena emerging from the model. As most models of planetary habitability focus on abiotic phenomena alone, future work should combine these abiotic models, with a biotic model such as ExoGaia to investigate the impact of adding biotic feedback. 

We will use real world language such as `planet' and `temperature' when discussing ExoGaia as the model is inspired by real world questions. However, ExoGaia is not intended to be a realistic model of planetary formation or dynamics; ExoGaia is an abstract model of a thought experiment, in line with e.g. Daisyworld (\cite{Watson1983}) that can be used to generate hypotheses about real planets - can a biosphere perform planetary regulation? Do Gaian bottlenecks occur?

This work builds on previous Gaian model research. There is a large body of literature on the Gaia hypothesis and on the many models used to investigate this hypothesis and so an in-depth review of Gaia will not be given here but the reader is pointed to  \cite{Boston1993,Lovelock2000,Schneider2013} for background on the Gaia  hypothesis, and \cite{Downing1999, Wood2008, McDonald-Gibson2008, Williams2010, Dyke2013, Nicholson2017, Arthur2017} for an overview of some key Gaian models investigated to date.

\section{The ExoGaia Model}

ExoGaia is heavily based on the `Flask' models \cite{Williams2007,Williams2008,Williams2010,Nicholson2017}, and shares similarities with the `Greenhouse world' model \cite{Worden2010,Worden2011}. We will first describe the model, then point out the key similarities and differences between these models and ExoGaia. An in depth model description is given in Appendix A.

\subsection{Model outline}

We model simple `planets' with well-mixed planetary atmospheres, the composition of which influences planetary temperatures. These planets are host to microbial life that consume and excrete atmospheric chemicals. All planets orbit a `star' that provides incoming radiation. We use the following terminology to describe ExoGaia:

\begin{itemize}
\item{Chemical} - a particular chemical species. Each chemical has either a cooling (e.g. reflective, high albedo) or warming (e.g. insulating) effect on the atmosphere.
\item{Chemical Set} - as the set of chemical species present in the system. 
\item{Geochemistry} - the static network of geochemical links between chemical species, i.e. the abiotic processes.
\item{Connectivity} - the probability of any two chemical species in a chemical set being connected by a geochemical process (also referred to as a link or connection). 
\item{Planet} - a system with a unique chemical set and geochemistry combination. 
\item{Biochemistry} - the biological links created by microbe metabolisms forming the biochemical network. Unlike the geochemistry of a planet, the biochemistry is not fixed; it evolves as a function of microbial evolutionary and ecological dynamics.
\item{Abiotic Temperature ($T_{abiotic}$)} - the temperature of a planet without life when its atmosphere is in equilibrium. Most of our simulated planets have abiotic temperatures that are inhospitable to life. (For the results presented in this paper the majority of simulated planets (over $70\%$) have inhospitably high abiotic temperatures. Appendix B explores alternative scenarios).
\end{itemize}

\begin{figure}[htbp!]
\centering
        \makebox[0.5\columnwidth][c]{\includegraphics[width=0.75\columnwidth]{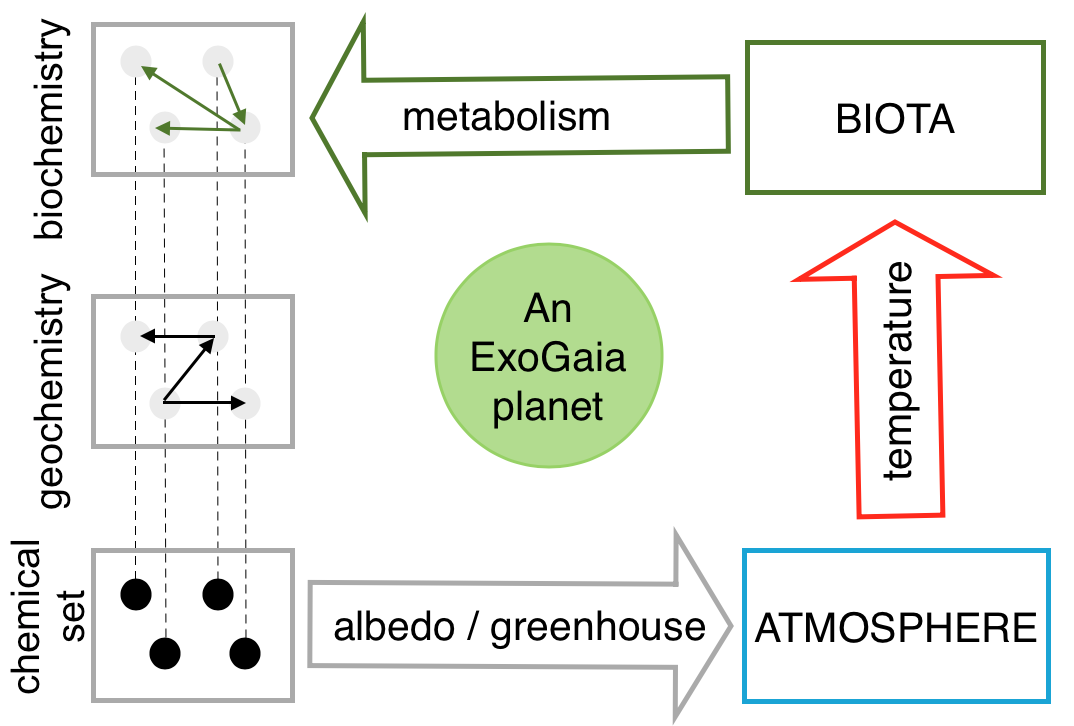}}
    \caption{The ExoGaia model schematic. Circles represent chemical species and arrows represent the geochemical or biochemical links between them.} \label{fig:1}
\end{figure}

Figure \ref{fig:1} shows a schematic for the ExoGaia model illustrating how each part of the planet - the chemistry, geochemistry and biochemistry are connected. We use agent based dynamics to model our ExoGaia experiments and therefore time is represented in model `timesteps'. 

\subsection{Microbes}

Model microbes consume and excrete atmospheric chemicals. Microbe metabolisms are genetically encoded and assume an external energy source, i.e. a star. The temperature of the host planet, $T_{planet}$, affects microbe metabolisms, and for simplicity all microbes share the same temperature preference, $T_{pref}$. At $T_{planet} = T_{pref}$ microbial growth rates will be at the maximum. As $T_{planet}$ moves away from $T_{pref}$ the microbes' consumption rate decreases and the growth rate drops. If the difference between $T_{pref}$  and $T_{planet}$ is too large, microbes will be unable to metabolise and will not consume/excrete any chemicals. Microbes die if their biomass drops below a certain threshold and there is also a constant probability of random death. If a microbe's  biomass reaches the reproduction threshold it reproduces asexually, with a constant probability of mutation for each gene, allowing new species to evolve on planets.

\subsection{Chemicals}

Model planet atmospheres are composed of various `chemical species'. There is a large body of literature on chemical reaction network theory \cite{Feinberg1987} which models the behaviour of real world systems and has been applied to planetary atmospheres, e.g. \cite{Sole2004}. We use a very simple chemical reaction network in ExoGaia.

Each chemical species has an insulating or a reflective property. We simplify real chemistry and limit a chemical species to being either insulating or reflecting, but not both. We can also take this simplification to be the overall impact a chemical has on the atmosphere. The collection of chemical species (and their greenhouse / albedo properties) possible on an ExoGaia planet is referred to as a `chemical set'. Not all chemical species in a chemical set might be present on a model planet. For a chemical species to be present it must be created by some process. The processes by which a chemical species can be created (or destroyed) are covered in later Sections on ``Atmosphere'', ``Geochemistry'' and ``Biochemistry''. 

All model atmospheric chemicals are assumed to be gaseous. Realistic atmospheric gases have both insulating and reflecting properties (via absorption and Rayleigh scattering) with the net effect depending on the abundance of the gas, the overall atmospheric mass \cite{Wordsworth:2013aa}, and the spectral energy of the host star \cite{Kaltenegger:2011aa}. In the ExoGaia model only the abundance of the gas determines it's overall impact on the host planet. In realistic scenarios, the outer edge of the Habitable zone depends on the limit where the condensation and scattering caused by adding more $CO_{2}$ to an atmosphere outweighs its greenhouse effect \cite{Kopparapu2013}.

\subsection{Temperature}
\label{subsection:2.4}

We use a linear approximation of the Stefan-Boltzmann law when calculating $T_{planet}$. This simplification has been shown to not greatly change the overall behaviour of the Daisyworld model \cite{Watson1983} \cite{Saunders:1994aa} \cite{De-Gregorio:1992aa} \cite{Weber:2001aa} \cite{Wood:2006aa}. The Stefan-Boltzmann equation is close to linear at real world habitable temperatures, i.e. near 22$^{o}$C. In ExoGaia, we are only interested in planetary dynamics when there is life on a planet, so while the `temperature' in the ExoGaia model is not constrained, we are only interested in a narrow range of habitable temperatures. The temperature behaviour outside this range is not important to the results. We will be using an unrealistic $T_{pref}$ for our model microbes to highlight the abstract nature of the model, however as a near linear relationship exists at habitable conditions on Earth, and we are striving to simplify the model abiotic environment as much as possible, we use a linearised Stefan-Boltzmann law in our model and take $\beta \propto T$, where $\beta$ is the energy provided to the planet by the host star per timestep and $T$ is temperature. We then make a further simplification and take the value of $\beta$ to be equal to the value of $T$. Appendix B4 further explores this temperature simplification.

\subsection{Atmosphere}

Many real planets have (or had), for example, volcanoes that spew forth aerosols and gases which come from the crust and the mantle. Gases can be lost from the planet's atmosphere by processes such as sublimation or some gases (e.g.  hydrogen) can be lost to space. We abstract these processes in the ExoGaia model. 

All model planets start with an `empty' atmosphere, and a constant inflow of chemicals from an external source begins at the start of each experiment. The `source chemicals' are the subset of chemical species in the chemical set that experience this inflow. Non-source chemicals do not exist on a planet unless created via a geochemical or biochemical process. There is a constant rate of atmospheric chemical outflow, performed by removing a fixed percentage of the well-mixed atmosphere each timestep. There is no spatial structure in the model.

A planet's atmospheric composition influences $T_{planet}$. We define $A_{I}$ as the fraction of the planet's current thermal energy retained by the atmosphere via insulation, and $A_{R}$ as the fraction of incoming radiation reflected by the atmosphere. Using the simplification described in Section \ref{subsection:2.4}, the value of $\beta_{planet}$ is the value of $T_{planet}$. Therefore $(1 - A_{I}) \beta_{planet}$ is equivalent to a planet's temperature decrease due to energy radiation into space, where $\beta_{planet}$ is the thermal energy of the planet. Similarly, $(1 - A_{R}) \beta_{star}$ is equivalent to the increase in temperature due to incoming solar radiation, where $\beta_{star}$ is the incoming solar radiation to a planet per timestep. Therefore a stable temperature is achieved if: 

\begin{equation}
\label{peq:1}
(1 - A_{I}) \beta_{planet} = (1 - A_{R}) \beta_{star}
\end{equation}

The values of $A_{I}$ and $A_{R}$ depend on the chemical composition of the atmosphere, and exist in the range $[0,1]$. This relation is described in an equation in Appendix A.
We calculate $\beta_{update}$, the updated thermal energy of a planet including the insulating and reflecting effect of the atmosphere, in the following way:

\begin{equation}
\label{peq:2}
\beta_{update} = A_{I} \beta_{planet} + (1 - A_{R})  \beta_{star}
\end{equation}

We neglect to model the complexities of atmospheric absorption in ExoGaia as that level of realism is unnecessary given the abstract simplified nature of the model. We also see that each timestep:

\begin{equation}
\beta_{lost} = (1 - A_{I}) \beta_{planet} + A_{R} \beta_{star}
\end{equation}

of energy is lost to space either as radiation from the planet or as reflected solar radiation. Although real stars age and change in luminosity, we keep our model simple and keep $\beta_{star}$ constant, to investigate the habitability of planets without external perturbation. This also makes sense biologically when considering the generation length of a microbe. It would take very, very many generations of microbes for a star to alter its solar radiation in a significant way. 

If $A_{I}  = 1$, a planet will perfectly insulate, and if $A_{R} = 1$ a planet will perfectly reflect all incoming radiation. Neither of these extremes is physically realistic; no atmosphere can perfectly insulate, nor reflect all incoming radiation, however this approach was favoured over choosing an arbitrary cut-off value. We are interested in regulation on habitable planets and in our experiments, the probability of $T_{planet}$ equalling $T_{stable}$, the temperature required for a stable microbe population, at these limits is extremely unlikely. Taking Equation \ref{peq:1}, if $A_{I} = 1$, then $A_{R} = 1$ must also be true for a constant $T_{planet}$. For long-term habitability, $A_{R} = A_{I} = 1$ must occur when $T_{planet} = T_{stable}$. This is highly unlikely and this scenario was not found to have happened in the results presented in this paper. Therefore this simplification does not impact on the conclusions drawn from our model.

\subsection{Geochemistry}

Geological links, or reactions, represent geological activity and take the form of $A \rightarrow B$, where $A$ and $B$ are different chemical species. This is a simplification of real chemistry where multiple reactants come together to form multiple products. Keeping the geological reactions simple allows us to more easily track chemicals through the system as they are converted via geological processes. 

\begin{figure}[htbp!]
\centering
    \begin{subfigure}{0.40\columnwidth}
        \centering
        \makebox[0.5\textwidth][c]{\includegraphics[width=1.0\textwidth]{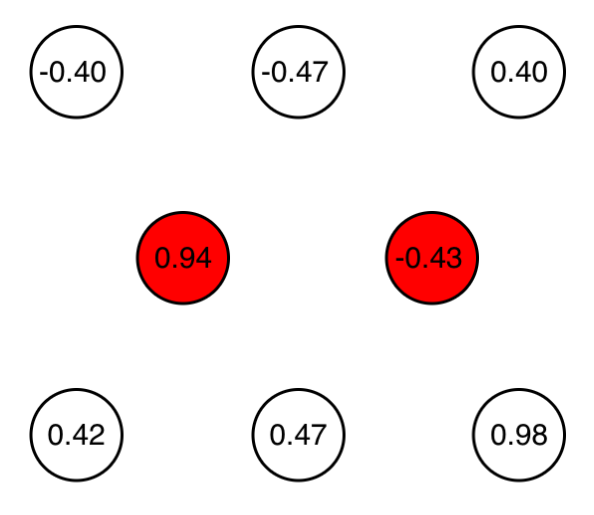}}
        \caption{Chemical set A \newline}\label{fig:2a}%
    \end{subfigure} %
    \hspace*{5mm}
    \begin{subfigure}{0.40\columnwidth}
        \centering
        \makebox[0.5\textwidth][c]{\includegraphics[width=1.0\textwidth]{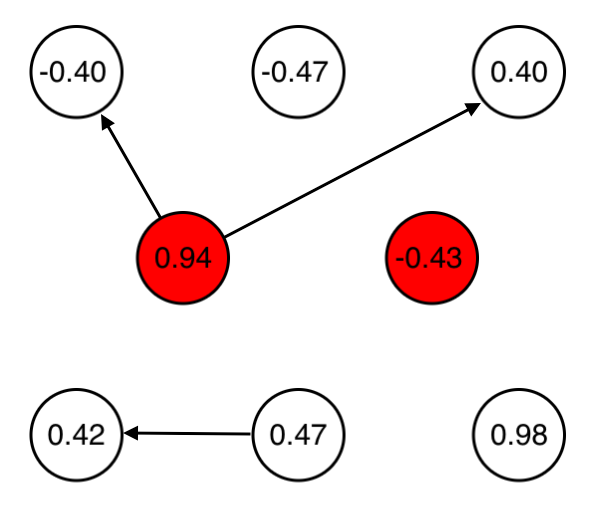}}
        \caption{Weakly connected \newline geochemistry}\label{fig:2b}%
    \end{subfigure} %
        \begin{subfigure}{0.40\columnwidth}
        \centering
        \makebox[0.5\textwidth][c]{\includegraphics[width=1.0\textwidth]{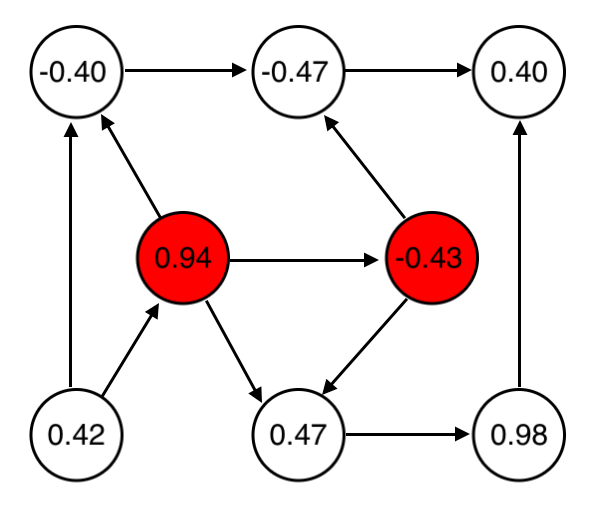}}
        \caption{Highly connected \newline geochemistry}\label{fig:2c}
    \end{subfigure} %
    \hspace*{5mm}
    \begin{subfigure}{0.40\columnwidth}
        \centering
        \makebox[0.5\textwidth][c]{\includegraphics[width=1.0\textwidth]{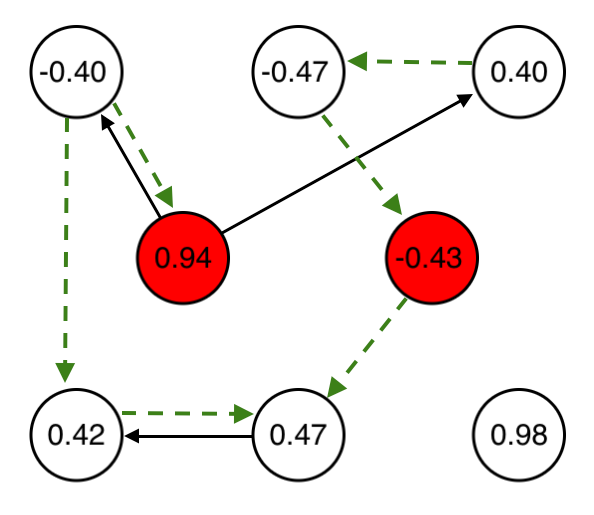}}
        \caption{Example biochemistry \newline}\label{fig:2d}%
    \end{subfigure} 
    \caption{Geochemical and biochemical networks. Circles represent chemical species. The number inside each chemical species is its greenhouse (positive) or albedo (negative) property. Red circles are source chemicals; they have an influx from outside the system. Black solid arrows represent the geochemical links between chemical species. Green dashed lines represent biochemical links caused by microbes' metabolisms. The size of the circle does not represent the abundance of a chemical species; if a chemical species is not a source chemical nor has any geochemical or biochemical processes producing it, it has an abundance of zero.} \label{fig:2}
\end{figure}

Geochemical links are generated based on a connectivity parameter $C$, which has a value between $[0, 1]$. $C = 0.2$ would determine a 20$\%$ probability for any pair of chemical species to be connected by a geochemical process. The direction of the link connection determines which  direction a process take  place, i.e.  $A \rightarrow B$  or  $B \rightarrow A$. We limit geological processes to acting in only one direction, i.e. if $A \rightarrow B$ then $B \rightarrow A$ is not allowed. We therefore describe only the net movement of chemicals linked by a geological process. The direction of a process has equal probability of acting in either direction. The link `strength' determines how strong a geological process / reaction is, and is taken from the range $[0,1]$. A link strength of $L = 0.3$ in the direction A  $\rightarrow$ B would mean that per timestep, 30$\%$ of the particles of chemical A would be converted into chemical B. Figure \ref{fig:2} depicts two different geochemistries. As chemical abundances change, the rate of a geological process will change. E.g. a geological process of the type A $\rightarrow$ B will happen at a faster rate when chemical A is abundant compared to when it is scarce. 

Geochemical links are not temperature dependant and remain constant throughout each experiment, therefore there are no geochemical temperature feedback loops in ExoGaia. Although many real world processes, e.g. silicate weathering, are temperature dependant, to isolate regulating effects caused by the microbes we remove this aspect from our model. This allows us to be confident that any regulation emerging in ExoGaia is due to the actions of the biosphere. This simplification does however limit the realism of the model and thus limit its applicability to real planets.

\subsection{Biochemistry}

Model microbes form temperature dependant biochemical links via their metabolisms, e.g. a species that consumes chemical A and excretes chemical B forms the biochemical link: $A  \rightarrow B$. The strength of a biochemical link depends on the number of microbe with the corresponding metabolism. Unlike the geochemical network, the biochemical network is not static; Biochemical links can change in strength, appear, and disappear, over time as the microbe community changes. Biochemical links can act in both directions, e.g. the biochemical links $A \rightarrow B$ and $B \rightarrow A$ are allowed to exist simultaneously. An example biochemistry is depicted in Figure \ref{fig:2}. These microbe metabolisms are highly simplified having only a single chemical reactant and single chemical product. Real microbe metabolism are more complex with multiple reactants and products. Using simplified microbe metabolisms allows for easier tracking of chemicals around ExoGaia systems, and makes the network diagrams presented later in this paper easier to produce and interpretable. Versions of the Flask model, on which ExoGaia is heavily based, have explored more complex microbe metabolisms with abiotic environmental regulation remaining a feature of these models \cite{Williams2008} \cite{Nicholson2017}.

The outflow of chemicals from the atmosphere is kept low, such that the timescale for a chemical to completely leave the atmosphere once produced by microbial activity is far longer than the typical lifespan of a microbe. This decouples the selection on individuals from their environmental effects and allows for long-term consequences (when compared to the average lifespan of a microbe) to occur from microbe activity. One real world example of this is the time it would take for our atmosphere on Earth to lose most of its $O_{2}$ if photosynthesis suddenly `switched off'. If a species evolves with a metabolism that produces a chemical not currently abundant in the atmosphere - $C_{new}$, a different species that consumes $C_{new}$ needs to emerge quickly before it builds up enough to disrupt the temperature regulation, or the species producing $C_{new}$ must die out, otherwise the whole community is susceptible to extinction.

\subsection{Planets}

We define a planet as a system with a particular chemical set and geochemistry. We can therefore run many experiments on a single planet to determine whether a planet has differing end states depending on early conditions.

No external forcing is present on our planets. Each planet's geochemistry remains fixed throughout an experiment and  the incoming  radiation $\beta_{star}$  remains  constant. Real planets are subjected to changing  host star luminosities and  changing  rates of geological processes over time, however  to understand how the biota  are able to adapt their host planet, we keep the environment fixed. It is then clearer when emerging phenomena are due to the biota.

An in-depth description of the model can be found in Appendix A. 

\subsection{New Features}

ExoGaia is based on the Flask model \cite{Williams2007,Williams2008,Williams2010,Nicholson2017}, which features model `flasks' containing microbe communities. These flasks experience an inflow and outflow of `nutrients', with the inflow medium at a constant `temperature'. Microbes change the temperature directly as a byproduct of their metabolism - increasing or decreasing it by a set amount. Differing from previous models such as Daisyworld \cite{Watson1983}, microbe's do not have localised space, however temperature regulation still robustly emerges. In the ExoGaia model, instead of microbes directly affecting a temperature, they impact $T_{planet}$ via consuming and excreting atmospheric chemicals. Also differing from the Flask model, the microbes are introduced to an ExoGaia planet, in most cases, before the atmosphere has reached equilibrium. 

`Greenhouse World' \cite{Worden2010, Worden2011} is a model of microbe communities interacting with insulating chemicals via their metabolism to regulate their environmental temperature. Although similar, ExoGaia has some key differences. Firstly, mutation only takes place in Greenhouse world when the system is in a stable state. Second, Greenhouse systems are seeded with a diverse community of microbes. These communities then reorganise via species dying off until a stable configuration is reached. Greenhouse world therefore demonstrates how diverse communities can scale down to a stable state, whereas in ExoGaia we seed with a single species, and the microbe community must evolve suitable metabolisms to regulate their environment, thus building up a regulating community where Greenhouse world reduces down. All life on Earth shares a common ancestor \cite{Sapp2009}, and so while it may theoretically be possible for life to form independently multiple times, that does not seem to be the case on Earth, and so we mirror this behaviour in our model.

A slow outflow of chemicals from a planet's atmosphere means that the consequences of microbial actions persist longer than their average lifespan - an important feature  not  present in previous models - allowing us to see how communities of microbes react to the long-term effects, especially the negative effects, of their metabolism. 

\section{Method}

Using this model, we investigate how the geochemical network of a planet affects the planet's colonisation success and the long-term  habitability.

We set the incoming radiation from the `star' per timestep $\beta_{star} = 500$ and set all microbes to share a preferred temperature $T_{pref} = 1000$. As this $T_{pref}$ corresponds, in our model, to a thermal energy of $\beta_{pref} = 1000$, we see that for a planet to reach habitable conditions, it must have an insulating atmosphere. Recall that all temperatures and energy values in the ExoGaia model are abstract. We generate the insulating / reflective properties of each chemical in our Chemical set by drawing a random number from the range $[-1, +1]$. A negative value means a chemical species is reflective, and a positive means it is insulating. We have 8 chemical species in our chemical set. We choose a chemical set such that the average effect of each chemical species is insulating. As $\beta_{star} < \beta_{pref}$, choosing an overall insulating chemical set insures many planets in our experiments will reach habitable planetary temperatures. This allows us to investigate how the microbe community interacts with it's host planet. Choosing an insulating chemical set does bias us to see more potentially habitable planets and thus increase the number of experiments where long-term habitability may be possible, but it does not help microbe communities, once seeded, in regulating their planet. The quantitative values produced by the ExoGaia model cannot be translated into real world values for an abstract model such as this. The qualitative behaviour of the model is the key point of interest. Chemical set A is used for the results presented in this paper, see Table \ref{table:1}.

\begin{table}
\centering
\caption{The greenhouse and albedo properties of chemical set A. The bold chemicals represent the source chemicals. The values in the table represent each chemical's impact on the atmosphere - insulating if positive, and reflective if negative.}\label{table:1}
 \begin{tabular}{cc}
    \hline
    Chemical & Greenhouse / albedo properly \\ \hline
    1 & -0.40 \\ 
    2 & -0.47 \\ 
    3 & 0.40 \\ 
    \textbf{4} & \textbf{0.94} \\ 
    \textbf{5} & \textbf{-0.43} \\ 
    6 & 0.42 \\ 
    7 & 0.47 \\ 
    8 & 0.98  \\ \hline
    Mean & 0.23 \\ \hline
     \end{tabular}
\end{table}

Despite sharing the same chemical set, planets vary hugely from one another due to their geochemical networks. These networks will determine how fast temperatures change, and the value of $T_{abiotic}$, for each planet. As we will see, sharing a chemical set does not result in identical planetary behaviours. The huge number of geochemical network configurations allows for many unique planets. In Appendix B, we present results from experiments with alternative chemical sets, but the same $\beta_{star} = 500$ value, exhibiting the same model behaviours presented with chemical set A, thus showing that chemical set A is not a unique case. 

We investigate a range of geological connectivity, $C$, for our planets: $ C = [0.1, 0.2, 0.3, 0.4, 0.5, 0.6, 0.7, 0.8, 0.9]$. As our model is abstract, we do not know what connectivities might be represented in the real world and so we cover almost the full range of possible values excluding $C = 0$, as we certainly live in a world of chemical reactions, and $C = 1$ as not every chemical can react with every other in real world chemistry. By exploring this large range we can investigate the effect connectivity has on the habitability of a planet and see how important this parameter is to the system dynamics.

We perform the following steps for each connectivity in list $C$:
\begin{enumerate}
\item Set up the planet's geological network
\begin{enumerate}
\item Begin the geological processes on the planet, allowing atmospheric chemicals to build up
\item Seed planet with a single species when $T_{planet} = T_{pref}$ 
\item if $T_{pref}$ is never reached, seed after $5 \times 10^{4}$ timesteps
\item The experiment ends $5 \times 10^{5}$ timesteps after seeding
\end{enumerate}
\item Repeat  step b) 100 times with different random  seeds initialising the microbes
\item Repeat  steps  (a)  to (c)  100 times  with  different  random  seeds  initialising  the planet's geological network
\end{enumerate}

There is evidence suggesting that life appeared on Earth as soon as conditions allowed \cite{Nisbet2001}. We treat our simple ExoGaia planets in a similar manner, seeding the planet when $T_{planet} = T_{pref}$ (if this happens at all, some planets will never reach $T_{pref}$). Because of the way temperature is determined in the model, planet temperatures might never exactly match $T_{pref}$, so to ensure seeding happens we determine a suitable `seeding window' $S_{w} = [T_{pref}, T_{pref}+50]$. Seeding can occur when planet matches any temperature in $S_{w}$ but seeding can only occur once. If a seed window has not been passed after $5\times 10^{4}$ timesteps then an seeding attempt is made once, and the model then continues as usual for $50\times 10^{4}$ timesteps.

This means that we will often be seeding the planets before the atmosphere is at equilibrium, and the $T_{abiotic}$ of a planet will often be far too hot for our microbe life to survive - effectively undergoing a highly simplified geologically induced greenhouse runaway. We therefore want to investigate whether the model microbes, with their simplified metabolisms, can take control of their host planet once they appear and keep the planet's temperature within habitable bounds. 

When we seed our planet with a single species, we seed with a species that consumes chemicals currently available on the planet. Any life with an unviable metabolism would very quickly die out. We could continually seed randomly until a species took hold on the planet, but predetermining that species we are seeding with could potentially survive (it has a food source) saves time.

\subsection{Habitability}

There are two types of habitability of interest to us:

\begin{itemize}
\item Colonisation success - what percentage of the time a planet is able to support life for $t_{survive} > 10^{3}$ timesteps.
\item Long-term habitability / survivability - what percentage of the time a planet is able to support life for the entire experiment duration: $t_{survive} = 5 \times 10^{5}$ timesteps.
\end{itemize}

The colonisation success indicates whether planetary conditions were suitable to support a self sustaining population for some time. $10^{3}$ timesteps is twice as long as the timescale for microbe death; therefore if the biosphere survives longer than $10^{3}$ timesteps, conditions must have allowed microbes to consume enough food to reproduce at a high enough rate to support a stable population. Long-term habitability measures the microbes ability, once they have successfully colonised a planet, to maintain habitable conditions for long time spans. 

Over a number of experiments, if a planet has high colonisation success but low late term habitability, it is a planet where life is usually able to colonise the planet and become established, but often fails to survive to the end of the experiment. If a planet has equal colonisation success and long-term habitability, it means that once life is established on a planet, it always survives the full experiment.

\section{Results}

In a highly abstract model such as ExoGaia, quantitative results cannot be applied directly to real world data, however exploring the qualitative behaviour of the model demonstrates how biosphere-environment coupled systems, such as the Earth and other inhabited planets, might emerge and under what circumstances. We find that on a diverse array of planets, life is able to `catch' the planetary atmospheric evolution of it's host planet and maintain habitable conditions. For the majority of ExoGaia planets, the $T_{abiotic}$ of the planet is highly inhospitable, yet we find many model planets hosting biospheres for long time spans. This demonstrates that model biospheres are capable of preventing planetary temperatures from reaching uninhabitable levels, and thus in principle, of regulating planetary temperatures.

\begin{figure}[htbp!]
\centering
    \begin{subfigure}{0.49\columnwidth}
        \centering
        \makebox[0.5\textwidth][c]{\includegraphics[width=1.0\textwidth]{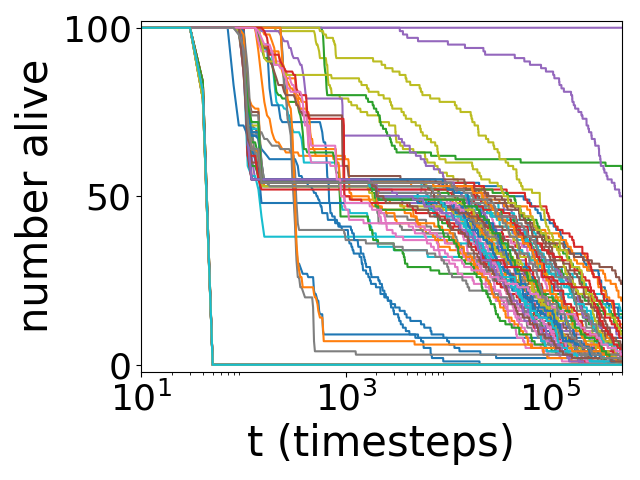}}
        \caption{C = 0.1}\label{fig:3a}%
    \end{subfigure} %
    \begin{subfigure}{0.49\columnwidth}
        \centering
        \makebox[0.5\textwidth][c]{\includegraphics[width=1.0\textwidth]{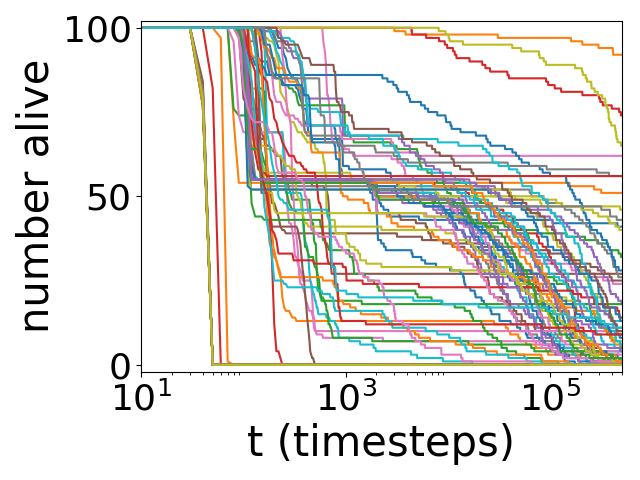}}
        \caption{C = 0.2}\label{fig:3b}%
    \end{subfigure} %
        \begin{subfigure}{0.49\columnwidth}
        \centering
        \makebox[0.5\textwidth][c]{\includegraphics[width=1.0\textwidth]{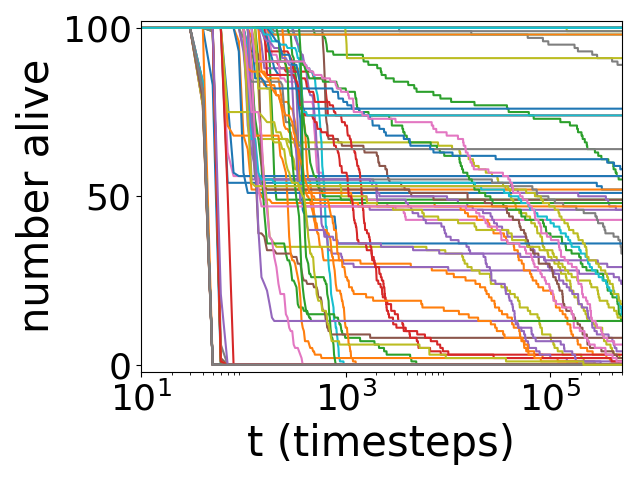}}
        \caption{C = 0.4}\label{fig:3c}%
    \end{subfigure} %
    \begin{subfigure}{0.49\columnwidth}
        \centering
        \makebox[0.5\textwidth][c]{\includegraphics[width=1.0\textwidth]{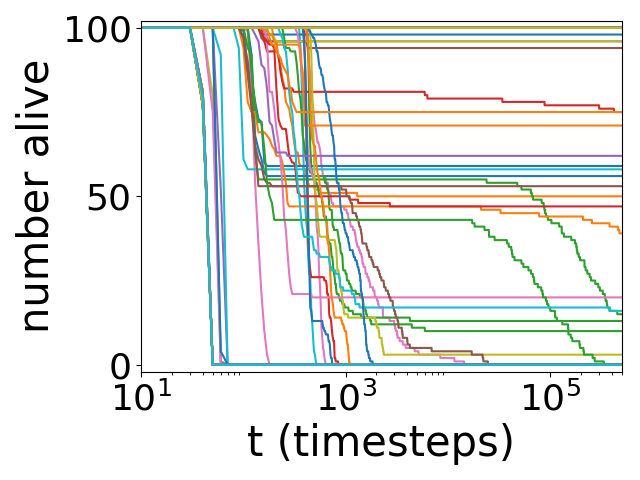}}
        \caption{C = 0.6}\label{fig:3d}%
    \end{subfigure} 
            \begin{subfigure}{0.49\columnwidth}
        \centering
        \makebox[0.5\textwidth][c]{\includegraphics[width=1.0\textwidth]{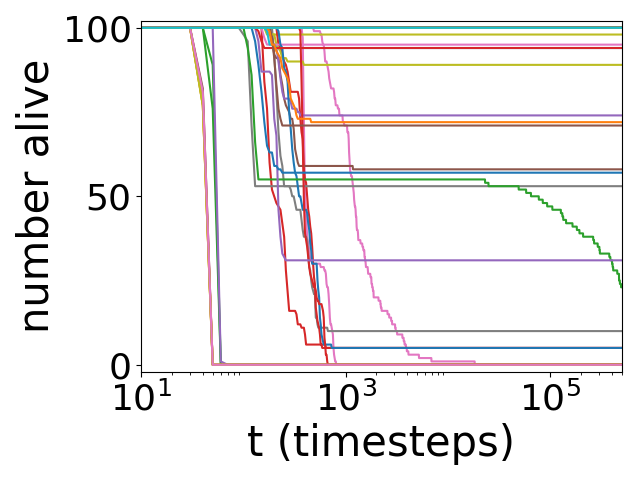}}
        \caption{C = 0.8}\label{fig:3e}%
    \end{subfigure} %
    \begin{subfigure}{0.49\columnwidth}
        \centering
        \makebox[0.5\textwidth][c]{\includegraphics[width=1.0\textwidth]{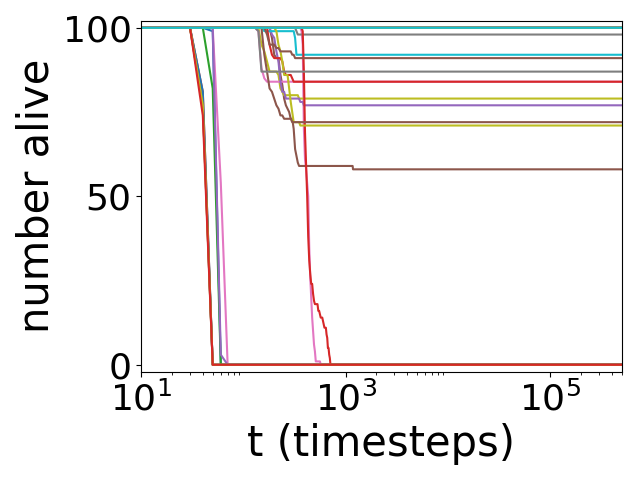}}
        \caption{C = 0.9}\label{fig:3f}%
    \end{subfigure} 
    \caption{Each line represents the survival curve of a planet. These survival curves tell us out of 100 experiments, how many are alive for each timestep. Critical planets show extinctions at random times, while Bottleneck planets show an early abrupt dying out and no deaths at longer timescales. Abiding planets have all 100 experiments alive for the whole experiment, and Extreme and Doomed planets always quickly die off early on in the experiment with non surviving to mid or long timescales. The first extinctions seen in each plot show the minimum time it takes for a microbe to starve to death. Note the log x-axis.} \label{fig:3}
\end{figure}

We find that colonisation success and long-term habitability success rates differ between model planets. As we performed 100 experiments on each planet, we can create a survival curve for each planet. Figure \ref{fig:3} shows the survival curves - the number of experiments (out of 100) with surviving life - for each planet against time (note the log x-axis). 

For low $C$, Figures \ref{fig:3a} and \ref{fig:3b}, there is no strong trend for when systems become extinct. Life is often able to successfully colonise a planet, but the planet is unlikely to experience long-term habitability. For higher $C$ we start to see planets with two distinct experiment outcomes: either life fails to colonise the planet and quickly goes extinct, or life successfully colonises the planet and survives the full experiment. For these planets, the colonisation success and the long-term habitability success of the planet are equal, meaning that if life is able to establish itself, it will survive for an indefinite period of time. This behaviour is explained in Section \ref{sec:bottleneck}. We see for $C = 0.9$, Figure \ref{fig:3f}, that all experiments either survive for the full duration, or become extinct early on, with no mid or late time extinctions taking place.

Table \ref{table:2.0} shows the number of planets that fail colonisation for all 100 experiments, and planets that had long-term habitability for all 100 experiments. In Figure \ref{fig:3} the number of planets that always immediately became extinct is difficult to determine, and it is not possible to see the number of planets that always survived the full experiment, so taking Figure \ref{fig:3} and Table \ref{table:2.0} together we get a more complete picture of the different planets' behaviour with changing connectivity.

\begin{table}
\centering
\caption{The number of planets that failed colonisation for 100$\%$ of experiments, and the number of planets that had long-term habitability (l.t.h.) success for 100$\%$ of experiments. The total number of planets simulated for each $C$ was 100.}\label{table:2.0}
 \begin{tabular}{ccc}
    \hline
    $C$ & 100$\%$ failed colonisation & 100$\%$ l.t.h success\\ \hline
    0.1 & 18 & 1 \\
    0.2 & 35 & 0 \\
    0.3 & 38 & 7 \\ 
    0.4 & 36 & 13 \\
    0.5 & 39 & 31 \\
    0.6 & 32 & 42 \\
    0.7 & 15 & 64 \\
    0.8 & 12 & 70 \\
    0.9 & 12 & 77  \\ \hline
     \end{tabular}
\end{table}

Based on our results we can determine 5 different classes of planet: \begin{itemize}
\item \textbf{Extreme} - Planets that never reach habitable temperatures
\item \textbf{Doomed} - Planets that do reach habitable temperatures but are unable to support life
\item \textbf{Critical} - Planets that can be successfully colonised by life, but go extinct at random times
\item \textbf{Bottleneck} - Life either fails to colonise these planets, or successfully colonises and enjoys long-term habitability - a bottleneck effect
\item \textbf{Abiding} - Life successfully colonises and experiences long-term habitability for all experiments
\end{itemize}

These planet class definitions are based only on two timescales: the colonisation success timescale which depends on the microbe death timescale; and the experiment length. 

We will now explain the regulation mechanism emergent in the ExoGaia model, and then show how a planet's geochemical network affects planetary habitability. We will then present example model planets to demonstrate various model behaviours, and finally show how planetary habitability is affected by connectivity.

\subsection{Regulation Mechanism}
\label{section:regulation}

The regulation mechanism takes the form of a negative feedback loop. All microbes share the same $T_{pref}$ and the same well-mixed environment, therefore any environmental change impacts all microbe species equally. There is no mechanism by which microbes can evolve only heating or cooling metabolisms, if abundant chemicals of any type are present on a system, microbes can, and will, evolve to consume those chemicals. Therefore it is the collective behaviour of the whole biosphere that leads to regulation rather than any specific microbe species. When $T_{planet} = T_{pref}$, assuming abundant chemicals, microbe populations will increase. The consumption rate of the microbes, $K$, drops as temperatures diverge from $T_{pref}$, therefore there are two temperatures where the value of $K$ will lead to a stable population: $T^{+}_{s} > T_{pref}$ and $T^{-}_{s} < T_{pref}$. For a stable $T_{planet}$ microbe populations must be stable.

When $\beta_{star} <  \beta_{pref}$, where $\beta_{pref}$ is the thermal energy of a planet at $T_{pref}$, an insulating atmosphere is required for habitable temperatures. This is the case for the results presented in the main body of this paper (alternative scenarios are explored in Appendix B). In this scenario, when $T_{planet} < T_{pref}$, the effects of increasing (+) $T_{planet}$ are: 

\begin{enumerate}
\item + $T_{planet}$ $\rightarrow$ + Population
\item + Population $\rightarrow$ - $T_{planet}$
\end{enumerate}

Flipping the signs, we also see that a decrease in $T_{planet}$ leads to an increase in $T_{planet}$. This forms a negative feedback loop. Increasing $T_{planet}$ improves habitability, which increases $K$, and thus increases microbe populations. This causes planetary cooling as the insulating power of the atmosphere is reduced via increased microbe consumption. Cooling degrades the environment, reducing microbe populations, and thus causes chemicals to build up in the atmosphere, increasing $T_{planet}$ and bringing us back to the start of the loop. This behaviour is known as `single rein-control' where the biota collectively form a single `rein' which `pulls' the system in one direction, while the abiotic processes on the planet `pull' the system in the other direction. Rein-control feedback mechanisms have been demonstrated in previous Gaian models, namely in Daisyworld \cite{Wood2008}, and the Flask model \cite{Nicholson2017}.

If instead $T_{planet} > T_{pref}$, the effects of increasing the temperature are now:

\begin{enumerate}
\item + $T_{planet}$ $\rightarrow$ - Population
\item - Population $\rightarrow$  + $T_{planet}$
\end{enumerate}

Now an increase in $T_{planet}$ degrades the environment for life and leads to further rises in $T_{planet}$ in a destabilising positive feedback loop. This results in microbe extinction. Temperature regulation therefore takes place at $T^{-}_{s}$ but not at $T^{+}_{s}$.  The behaviour seen in both feedback loops is known as feedback on growth \cite{Lenton1998}.

\begin{figure}[htbp!]
\centering
    \begin{subfigure}{0.49\columnwidth}
        \centering
        \makebox[0.9\textwidth][c]{\includegraphics[width=1.0\columnwidth]{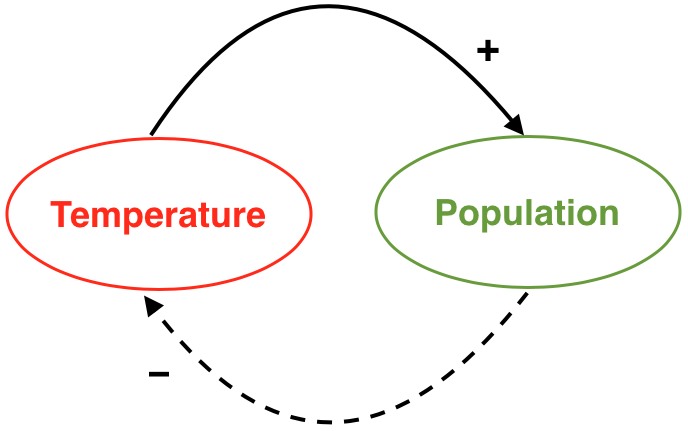}}
        \caption{Negative feedback loop for $T_{planet} < T_{pref}$}\label{fig:4a}%
    \end{subfigure} %
    \begin{subfigure}{0.49\columnwidth}
        \centering
        \makebox[0.9\textwidth][c]{\includegraphics[width=1.0\columnwidth]{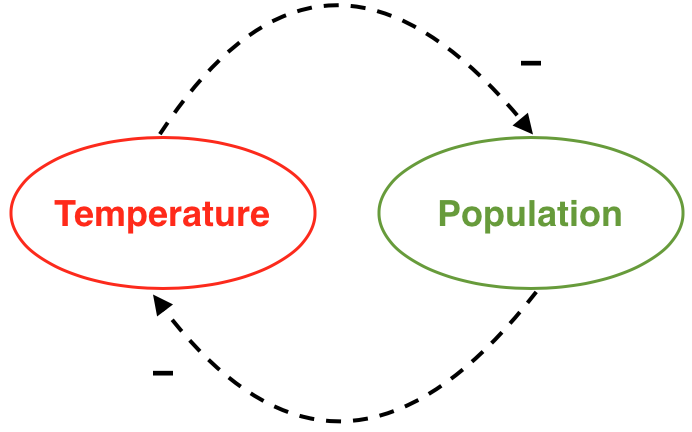}}
        \caption{Positive feedback loop for $T_{planet} > T_{pref}$}\label{fig:4b}%
    \end{subfigure} %
    \caption{The regulating negative feedback loop (a), and the destabilising positive feedback loop (b). Arrows indicate the effect of an increase in the source on the target. The sign indicates whether an increase in the source leads to an increase or decrease the target. In (a) an increase in temperature causes an increase in population, whereas an increase in population causes a decrease in temperature. This forms a negative feedback loop. In (b) an increase in temperature decreases the population, and a decrease in population further increases the temperature. This forms a positive feedback loop. } \label{fig:4}
\end{figure}

When $T_{planet} > T_{pref}$ a positive feedback loop in the opposite direction is possible, with runaway planetary cooling occurring until $T_{planet} < T_{pref}$, where the negative feedback loop takes over. However as $T_{abiotic} > T_{pref}$ for a habitable planet, when $T_{planet} > T_{pref}$ a reduction in temperature is unlikely; when habitability is low the abiotic processes on the planet dominate. If $T_{planet}$ rises to above $T_{pref}$, extinction is the expected outcome. Figure \ref{fig:4} shows the positive and negative feedback loops for $T_{planet} < T_{pref}$ and $T_{planet} > T_{pref}$.

\subsection{Geochemistry and Habitability}

We investigated the underlying geochemical networks for planets of each class to determine what lead to the different planetary behaviours, and found that a planet's geochemical network strongly determines its chance for long-term habitability success. We found two key properties: 

\begin{itemize}
\item The geochemical network must be such that planetary temperatures recover faster from any microbe induced cooling than the time it would take for the population to go extinct due to starvation.
\item For long-term habitability success, the geochemical network must provide many recycling chemical loops.
\end{itemize}

Different geochemical networks will lead to temperature changes taking place at different rates on different planets. As seen in Section \ref{section:regulation}, for potentially habitable planets, microbe populations cause planetary cooling. For a planet to be habitable, the geochemical network must be such that $T_{planet}$ increases after microbe induced cooling fast enough to avoid microbe extinction. The rate of temperature change due to abiotic processes alone plays a strong role in determining the colonisation success of a planet.

This is not enough to guarantee long-term habitability however. As seen in Figure \ref{fig:3}, many planets that were successfully colonised later went extinct. Planets that experienced long-term habitability all shared the feature of having a geochemical network that provided many chemical recycling loops. For an example, assume that there are only four chemicals in the chemical set and take the geochemical network $1 \rightarrow 2 \rightarrow 3 \rightarrow 4 \rightarrow 1$, where numbers represent chemical species and arrows are geological processes. In this example, for any microbe metabolism, the geochemistry recycles the waste product back to the food source. This allows a microbe community to `control' the entire atmosphere with only a single metabolism. By influencing the abundance of one chemical species in the loop, all other chemical species are impacted. Temperature regulation takes place in ExoGaia via the collective actions of the biosphere consuming the atmospherical chemicals without bias, therefore if there are many geochemical recycling networks, and microbes can influence the abundance of many chemical species with fewer metabolisms, achieving planetary regulation is likelier.

\begin{figure}[htbp!]
\centering
        \makebox[0.3\columnwidth][c]{\includegraphics[width=0.7\columnwidth]{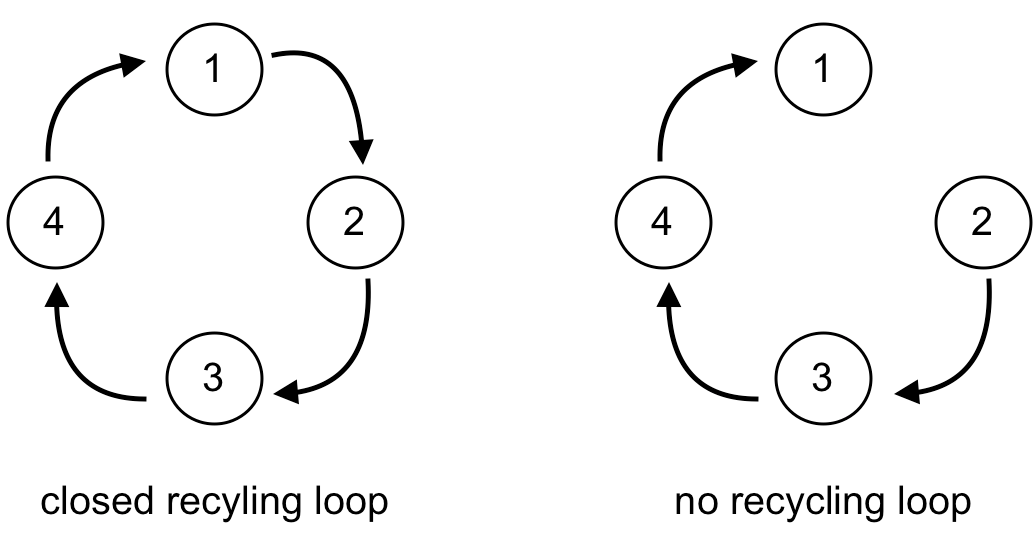}}
    \caption{Circles represent chemical species and arrows represent the geochemical links between them.} \label{fig:5}
\end{figure}

Now consider the geochemical network $2 \rightarrow 3 \rightarrow 4 \rightarrow 1$. Chemicals now accumulate as chemical species $1$, and the geochemical network does not recycle waste back to food for many metabolisms e.g. $2 \rightarrow 3$ or $3 \rightarrow 1$. These scenarios are depicted in Figure \ref{fig:5}. Biological links are temperature dependant and change as planetary conditions change. This makes them less stable than the temperature independent geochemical links. Therefore if a geochemical network does not have many recycling loops, and biology must `complete' many missing links, the system will be more sensitive to temperature changes. Biological links can amplify perturbations throughout the system as $T_{planet}$ impacts the biosphere, which impacts the biochemical network, which further impacts $T_{planet}$. Therefore these systems are highly susceptible to perturbation, and as any large-scale changes in temperature carry a risk of extinction, these systems are less likely to experience long-term habitability. 

\subsection{Example planets}

We now present an example planet for each planet class (each example planet has connectivity $C = 0.4$) to demonstrate how the underlying geochemical network impacts a planet's colonisation success and long-term habitability. 

\subsubsection{Uninhabitable planets}
\label{subsubsec:Dead}

The majority of model planets that fail in every experiment to support life have a $T_{abiotic}$ that's too cold for life to survive. Once seeded, life either cannot metabolise at all, or can only metabolise at levels too low for a stable population, leading to extinction. The underlying geochemistry doesn't have much effect here other than to convert the heating chemical species to cooling ones thus rendering the planet uninhabitable. We will refer to this type of uninhabitable planet as `Extreme' planets - planets with temperatures that never reach habitable levels.

A small number of uninhabitable planets have a $T_{abiotic}$ such that $T_{abiotic} \geq T_{pref}$. These planets typically have only weakly insulating atmospheres, and temperatures rise very slowly. On these planets when life is seeded, it consumes this insulating atmosphere and causes planetary cooling pushing the planet to uninhabitable temperatures. This in turn causes the microbe population to decline. With a smaller population, the abiotic processes on the planet dominate, however $T_{planet}$ does not rise to the bounds of habitability fast enough and life on the planet goes extinct. We refer to these planets as `Doomed' planets; although temperatures on these planets do reach habitable levels and microbes can initially metabolise, life always fails to colonise the planet.

\begin{figure}[htbp!]
\centering
    \begin{subfigure}{0.49\columnwidth}
        \centering
        \makebox[0.5\textwidth][c]{\includegraphics[width=1.0\columnwidth]{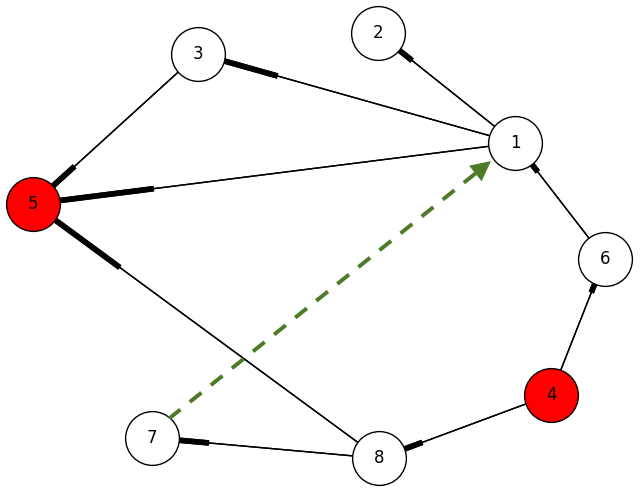}}
       \caption{$t_{seed}$} \label{fig:6a}
        \hspace*{5mm}
    \end{subfigure} 
        \begin{subfigure}{0.49\columnwidth}
        \centering
        \makebox[0.5\textwidth][c]{\includegraphics[width=1.0\columnwidth]{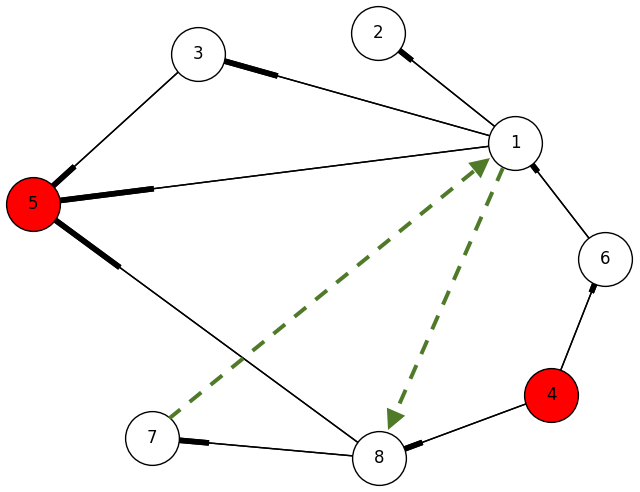}}
        \caption{$t_{seed} + 100$}\label{fig:6b}
        \hspace*{5mm}
    \end{subfigure} 
     \begin{subfigure}{0.49\columnwidth}
        \centering
        \makebox[0.5\textwidth][c]{\includegraphics[width=1.0\columnwidth]{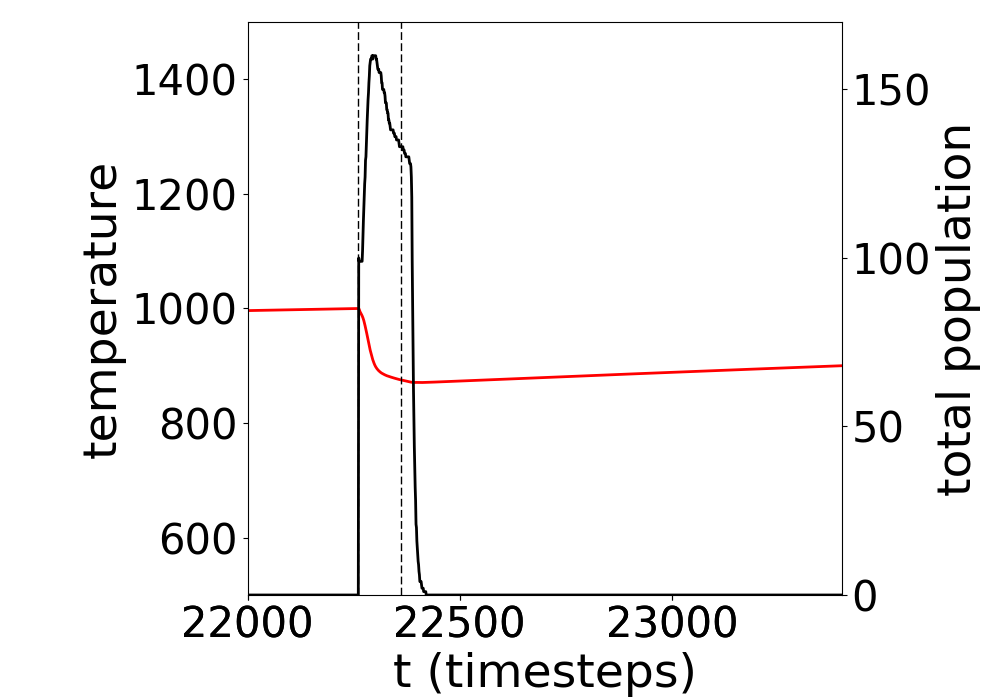}}
        \caption{Temperature and total population for early time}\label{fig:6c}
        \hspace*{5mm}
    \end{subfigure} 
\caption{\textbf{Example Doomed Planet}: Snapshots of two experiments for the same planet showing the geochemical network in black solid lines, and the biochemical network in green dashed lines. Red circles represent source chemicals. $t_{seed}$ is the time the planet was seeded with life. Plot c) shows temperature (red) and total population (black) against time. C = 0.4. The thick end of the geochemical links indicates positive direction.}
\label{fig:6}
\end{figure}

Figure \ref{fig:6} shows snapshots of the geochemistry and biochemistry of an example Doomed planet that had an $T_{abiotic}$ such that $T_{abiotic} > T_{pref}$. The static geochemistry is represented by black solid lines (with the thick end indicating a positive direction of chemical flow) and the non-static biochemistry is represented by green dashed lines and changes as the microbe community changes. Circles indicate chemical species with source chemicals as red circles. We see the microbe seeding occur when $T_{planet} = T_{pref}$. Microbes are able to establish biochemical links beyond the seed species (Figure \ref{fig:6b}), however the planet becomes extinct soon after. Figure \ref{fig:6c} shows that the planet's temperature was increasing very slowly before microbe seeding, and that planetary temperatures do not recover fast enough from microbe induced cooling to avoid microbial extinction. For this planet, the geochemical network was arranged such that abiotic temperature changes happen too slowly to counteract microbial cooling making the planet unsuitable for life. This behaviour, where life reduces the habitability of its environment, is often called `anti-Gaian' behaviour in contrast to `Gaian' behaviour where life enhances its environment's habitability.

This behaviour highlights an important feature of the model - a habitable temperature is not enough for habitability. All life interacts with its environment, removing and producing chemicals during metabolisation. As such, life requires an environment where interacting with the environment does not destroy habitability. On these `Doomed' planets, the atmosphere is only weakly insulating, and atmospheric depletion by the seeded microbes' consumption quickly results in inhospitably cold temperatures. As all life in this model experiences the same environment, it is not possible for microbes to evolve only metabolisms that consume cooling chemicals. If life cannot interact with its environment without pushing it past the bounds of habitability, then despite reaching habitable temperatures, such planets are not good candidates for hosting life. The behaviour of these planets when `reseeding' - life is reintroduced after going extinct - is included in the experiments is explored in Appendix B.

\subsubsection{Critical planets}

Critical planets often have high colonisation success however long-term habitability is unlikely. There is no obvious trend in when a Critical planet will become extinct. Critical planets tend to have geochemical networks that cause $T_{planet}$ to rise faster than seen in Doomed planets, meaning that Critical planet temperatures can recover from microbe induced cooling fast enough to prevent extinction. This provides a good environment for colonisation success, however, Critical planet geochemical networks do not provide a large number of chemical recycling networks, therefore certain chemical species can quickly accumulate in abundance and require microbe intervention to prevent large temperature changes.

\begin{figure}[htbp!]
\centering
    \begin{subfigure}{0.49\columnwidth}
        \centering
        \makebox[0.5\textwidth][c]{\includegraphics[width=1.0\columnwidth]{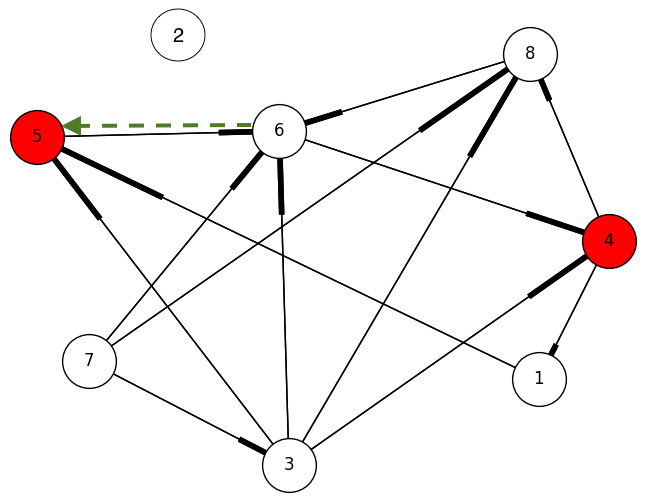}}
       \caption{$t_{seed}$} \label{fig:7a}
        \hspace*{5mm}
    \end{subfigure} 
        \begin{subfigure}{0.49\columnwidth}
        \centering
        \makebox[0.5\textwidth][c]{\includegraphics[width=1.0\columnwidth]{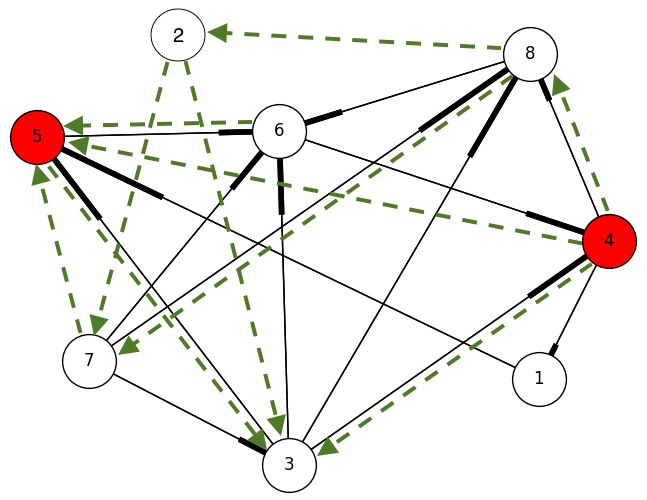}}
        \caption{$t_{seed} + 100$}\label{fig:7b}
        \hspace*{5mm}
    \end{subfigure} 
            \begin{subfigure}{0.49\columnwidth}
        \centering
        \makebox[0.5\textwidth][c]{\includegraphics[width=1.0\columnwidth]{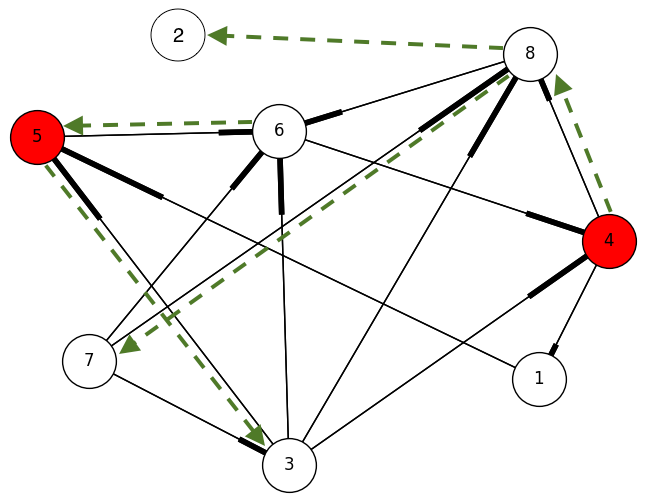}}
        \caption{$t_{seed}+500$}\label{fig:7c}
        \hspace*{5mm}
    \end{subfigure} 
            \begin{subfigure}{0.49\columnwidth}
        \centering
        \makebox[0.5\textwidth][c]{\includegraphics[width=1.0\columnwidth]{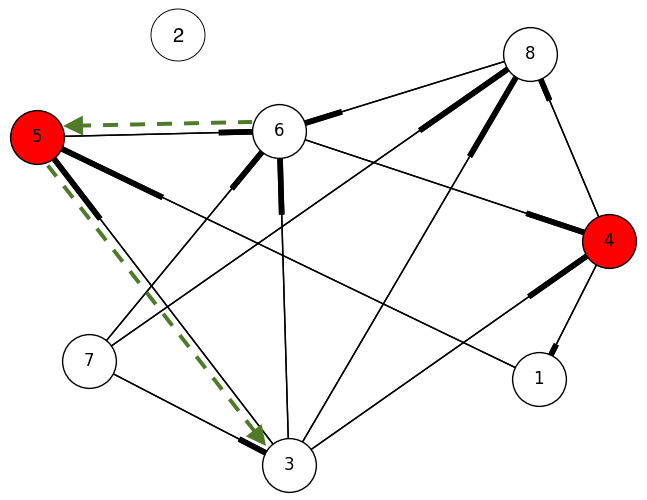}}
        \caption{$t_{seed}+1000$}\label{fig:7d}
        \hspace*{5mm}
    \end{subfigure} 
            \begin{subfigure}{0.49\columnwidth}
        \centering
        \makebox[0.5\textwidth][c]{\includegraphics[width=1.0\columnwidth]{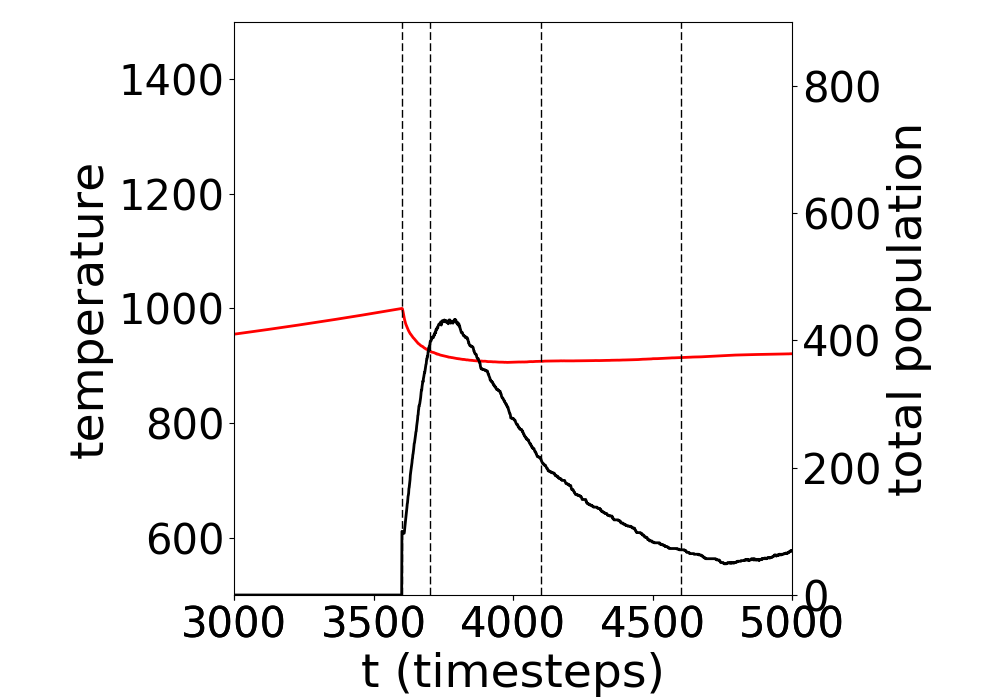}}
        \caption{Temperature and total population for early time\newline}\label{fig:7e}
        \hspace*{5mm}
    \end{subfigure} 
            \begin{subfigure}{0.49\columnwidth}
        \centering
        \makebox[0.5\textwidth][c]{\includegraphics[width=1.0\columnwidth]{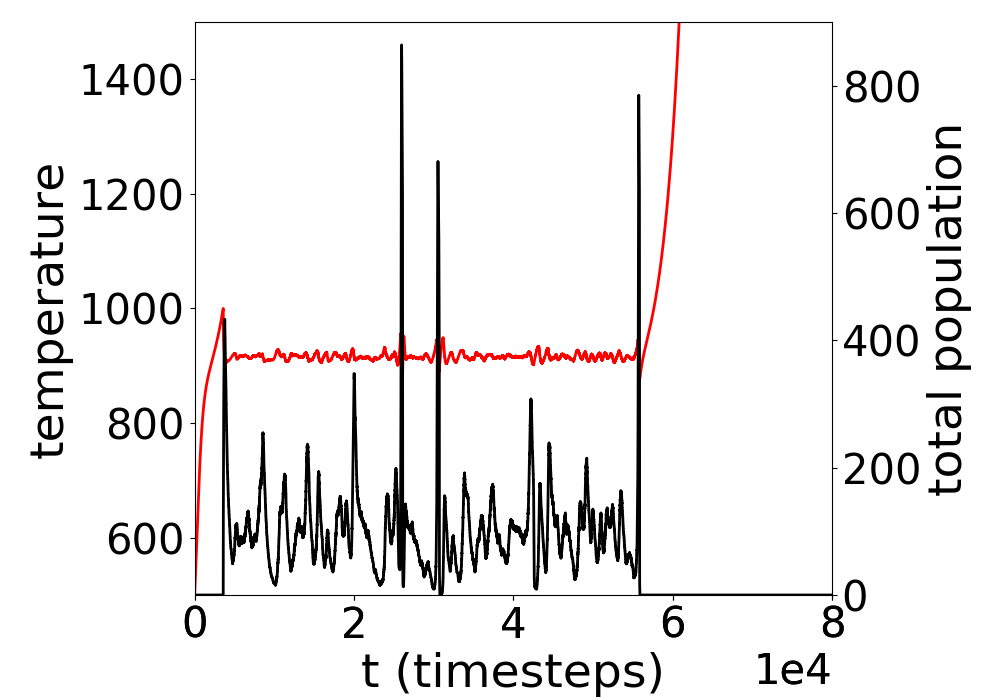}}
        \caption{Temperature and total population for the inhabited period of the experiment}\label{fig:7f}
        \hspace*{5mm}
    \end{subfigure} 
\caption{\textbf{Example Critical Planet}: Snapshots of a single experiment showing the geochemical network in black solid lines, and the biochemical network in green dashed lines. Red circles represent source chemicals. $t_{seed}$ is the time the planet was seeded with life. Plots e) and f) show temperature (red) and total population (black) against time. C = 0.4. The thick end of the geochemical links indicates positive direction.}
\label{fig:7}
\end{figure}

Figure \ref{fig:7} shows snapshots of the geochemistry and biochemistry on a Critical planet, with the temperature and population curves against time. We see that the biochemistry acts erratically; biochemical links quickly infiltrate the geochemical network but later disappear. Figure \ref{fig:7e} shows a large population spike after seeding which then dies down. Differing from the Doomed planet (Figure \ref{fig:6}), the temperature recovers fast enough from microbe induced cooling to avoid extinction, and the planetary temperature is then regulated by the microbes for approximately 55,000 timesteps, Figure \ref{fig:7f}. For Doomed planets, cooling by microbes results in extinction, however for this Critical planet, cooling prevents $T_{planet}$ from rising to inhospitable levels, and thus avoids microbial extinction. In Figure \ref{fig:7f} we can see the purely abiotic temperature behaviour of this planet when life goes extinct; we see that the planet's temperature immediately and rapidly climbs after microbial extinction. This demonstrates how the same behaviour by life could be classed as `Gaian' or `anti-Gaian' depending on the external environment.

Figure \ref{fig:7f} shows the total population fluctuates around a value of $\approx120$ with extreme population spikes happening a few times - the last of these causing the extinction of the system. These extreme population spikes occur due to the disconnected nature of the geochemical network; chemical species 2 is entirely unconnected to other chemical species geochemically. Figure \ref{fig:7c} show a biochemical link from $8 \rightarrow 2$, however no biochemical link converting chemical species 2 to any other chemical and thus the abundance of chemical species $2$ increases rapidly. If a microbe evolves that consumes chemical 2, it will have an abundant source of food. As chemical 2 is a cooling chemical (see Table \ref{table:2.0}), depleting this chemical species will heat the system, pushing $T_{planet}$ closer to $T_{pref}$ and increasing all microbes' reproduction rates, causing an explosion in population. This population explosion will cause overall depletion of the atmospheric chemicals, and thus, as on average the chemicals in Chemistry A are greenhouse chemicals, the temperature will cool and the population will die back down. This scenario is the cause of the first extreme spike seen in Figure \ref{fig:7f}. Not all Critical planets have completely unconnected chemical species as in Figure \ref{fig:7} but they share the common characteristic of a more disconnected geochemical network with fewer purely geochemical recycling loops. Biochemical links are more susceptible to oscillation as changes in one link can have knock on effect to others amplifying the perturbation, thus the more biochemical links required to close recycling loops, the less stable the system is. This is what makes Critical planets susceptible to total extinction.

\subsubsection{Bottleneck planets}\label{sec:bottleneck}

Bottleneck planets either fail to be successfully colonised, or are successfully colonised and life survives the full experiment. Bottleneck planets once successfully colonised are not susceptible to extinction.

\begin{figure}[htbp!]
\centering
    \begin{subfigure}{0.49\columnwidth}
        \centering
        \makebox[0.5\textwidth][c]{\includegraphics[width=1.0\textwidth]{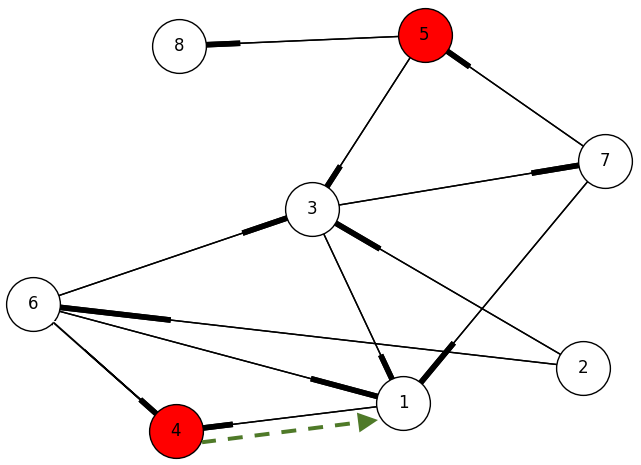}}
       \caption{$t_{seed}$} \label{fig:8a}
        \hspace*{5mm}
    \end{subfigure} 
        \begin{subfigure}{0.49\columnwidth}
        \centering
        \makebox[0.5\textwidth][c]{\includegraphics[width=1.0\textwidth]{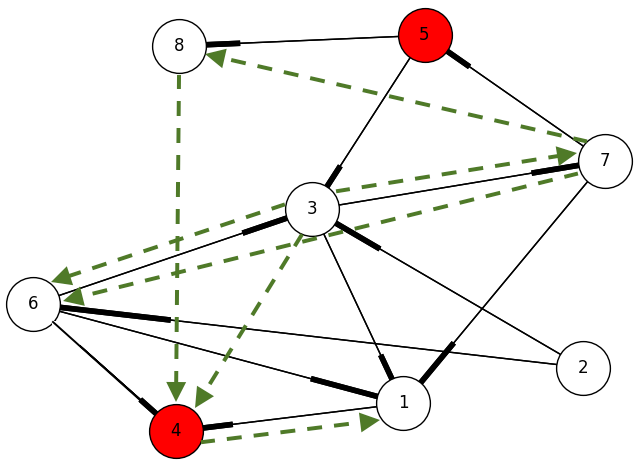}}
        \caption{$t_{seed} + 100$}\label{fig:8b}
        \hspace*{5mm}
    \end{subfigure} 
            \begin{subfigure}{0.49\columnwidth}
        \centering
        \makebox[0.5\textwidth][c]{\includegraphics[width=1.0\textwidth]{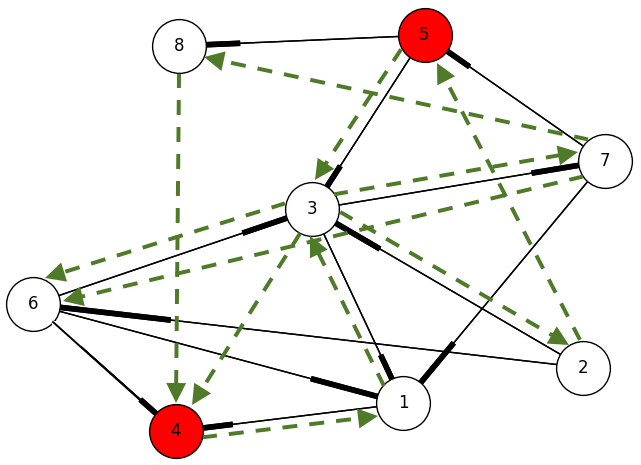}}
        \caption{$t_{seed}+500$}\label{fig:8c}
        \hspace*{5mm}
    \end{subfigure} 
            \begin{subfigure}{0.49\columnwidth}
        \centering
        \makebox[0.5\textwidth][c]{\includegraphics[width=1.0\textwidth]{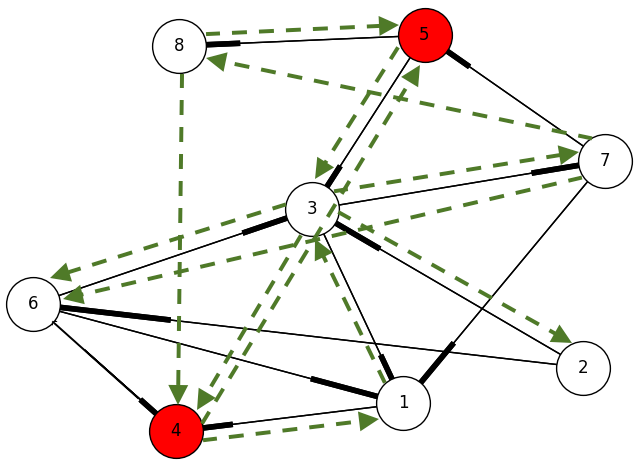}}
        \caption{$t_{seed}+1000$}\label{fig:8d}
        \hspace*{5mm}
    \end{subfigure} 
            \begin{subfigure}{0.49\columnwidth}
        \centering
        \makebox[0.5\textwidth][c]{\includegraphics[width=1.0\textwidth]{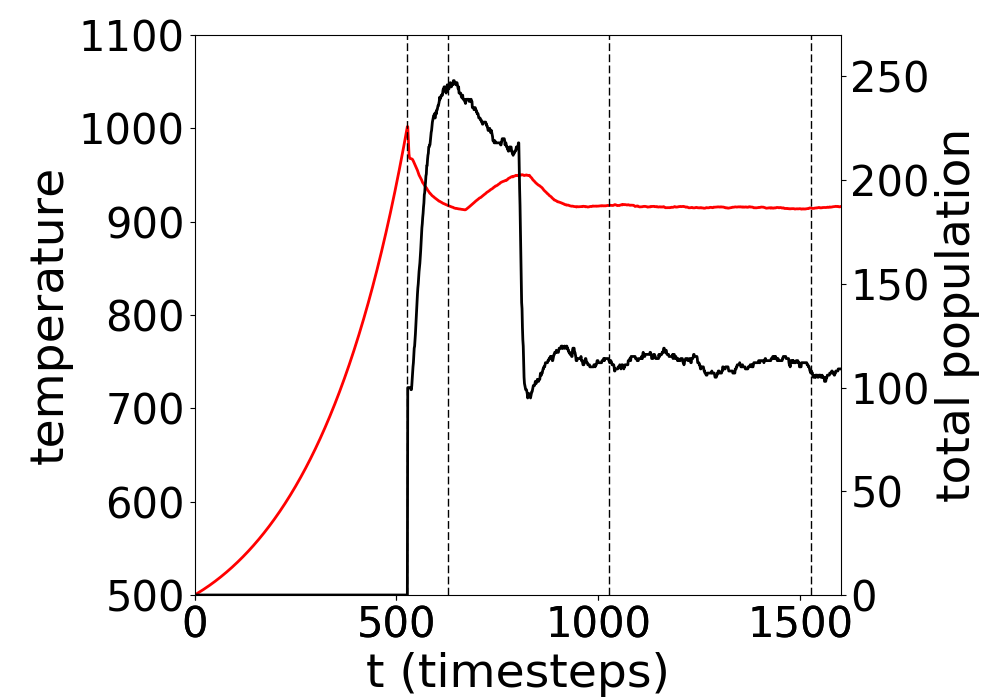}}
        \caption{Temperature and total population for early time\newline}\label{fig:8e}
        \hspace*{5mm}
    \end{subfigure} 
            \begin{subfigure}{0.49\columnwidth}
        \centering
        \makebox[0.5\textwidth][c]{\includegraphics[width=1.0\textwidth]{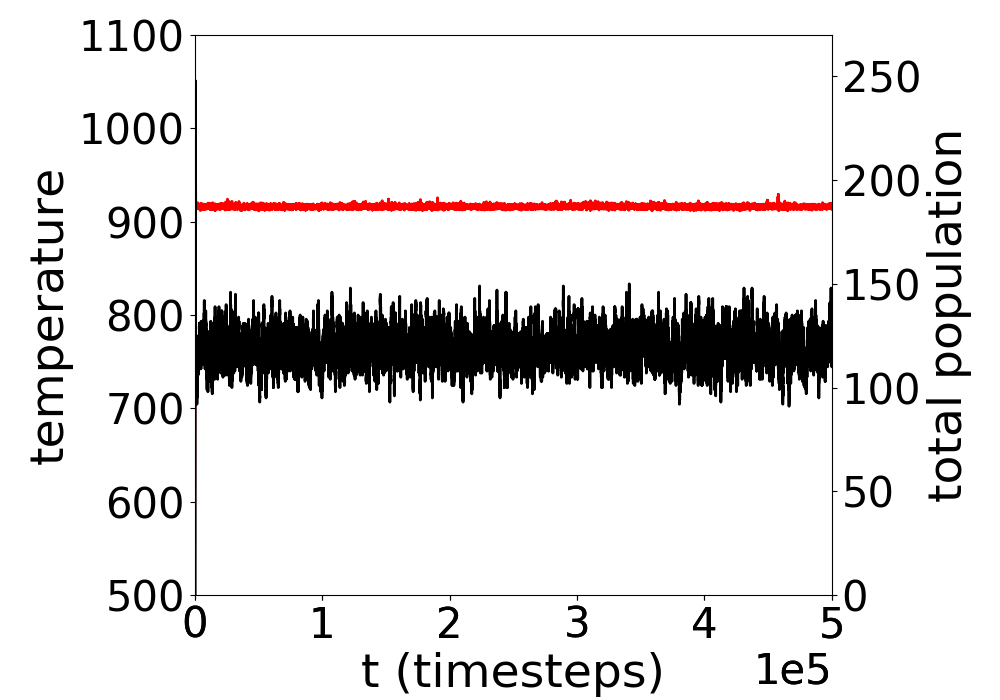}}
        \caption{Temperature and total population for full \newline experiment}\label{fig:8f}
        \hspace*{5mm}
    \end{subfigure} 
\caption{\textbf{Example Bottleneck Planet}: Snapshots of a single experiment showing the geochemical network in black solid lines, and the biochemical network in green dashed lines. Red circles represent source chemicals. $t_{seed}$ is the time the planet was seeded with life. Plots e) and f) show temperature (red) and total population (black) against time. C = 0.4. The thick end of the geochemical links indicates positive direction.}
\label{fig:8}
\end{figure}

Figure \ref{fig:8} shows snapshots of the biochemistry overlaid on the geochemistry for a Bottleneck planet. Examining the geochemistry we see that there are two chemical species, 8 and 4, with no geochemical process converting them to another chemical species. The initial seed species consumes chemical 4. After seeding, there is a population explosion and many new biochemical links are formed including metabolisms consuming 8. The system now has metabolisms controlling both these important chemical species. The population explosion and subsequent consumption of the atmospheric chemicals has caused $T_{planet}$ to cool, causing a sharp decline in the total population, allowing the abiotic processes to take over, warming the planet once more. This improves conditions for life allowing the population to rise again, this time to a more sustainable level, and $T_{planet}$ stabilises under the microbes' regulation. We see that there are many recycling loops already provided by the geochemistry, any waste (barring waste of chemical species 8 or 4) produced by a microbe can be recycled back into its food source, although some loops take more geochemical reactions than others. This makes it easier for the microbes to retain control over their planet's atmosphere as geochemical links, unlike biochemical links, are not prone to temperature dependant fluctuations.

\begin{figure}[htbp!]
\centering
    \begin{subfigure}{0.49\columnwidth}
        \centering
        \makebox[0.5\textwidth][c]{\includegraphics[width=1.0\textwidth]{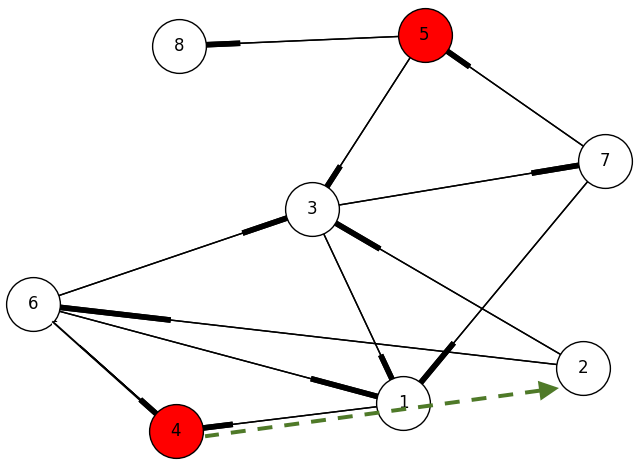}}
       \caption{$t_{seed}$} \label{fig:9a}
        \hspace*{5mm}
    \end{subfigure} 
        \begin{subfigure}{0.49\columnwidth}
        \centering
        \makebox[0.5\textwidth][c]{\includegraphics[width=1.0\textwidth]{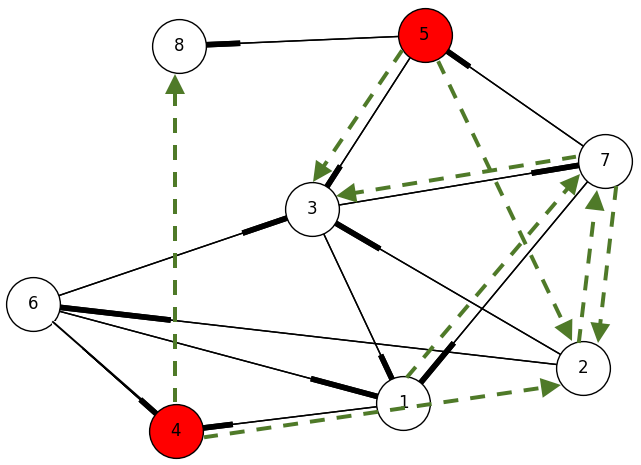}}
        \caption{$t_{seed} + 100$}\label{fig:9b}
        \hspace*{5mm}
    \end{subfigure} 
    \begin{subfigure}{0.49\columnwidth}
        \centering
        \makebox[0.5\textwidth][c]{\includegraphics[width=1.0\textwidth]{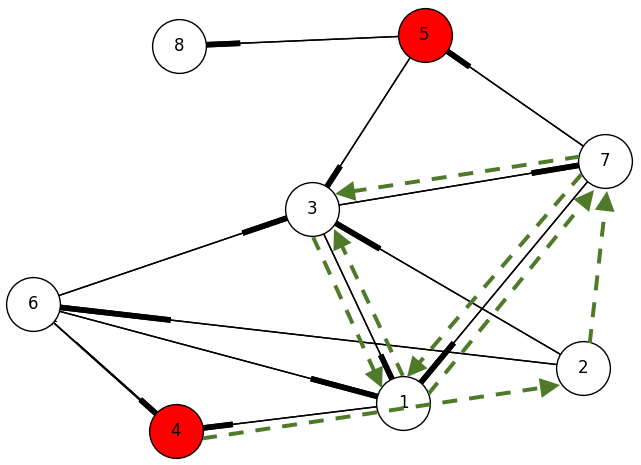}}
        \caption{$t_{seed} + 500$ \newline}\label{fig:9c}
        \hspace*{5mm}
    \end{subfigure} 
            \begin{subfigure}{0.49\columnwidth}
        \centering
        \makebox[0.5\textwidth][c]{\includegraphics[width=1.0\textwidth]{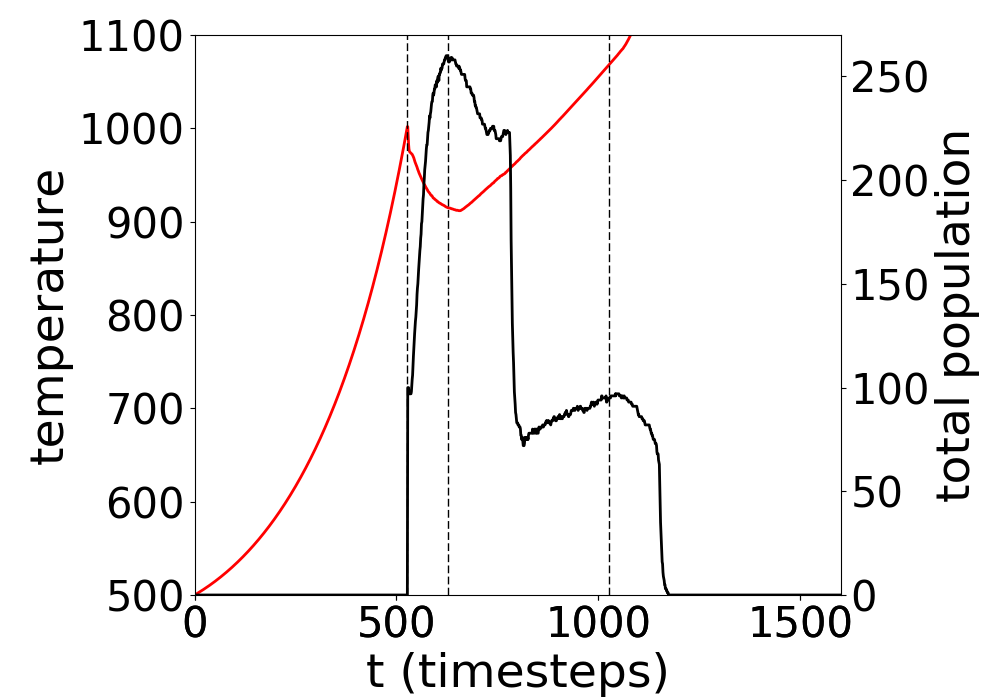}}
        \caption{Temperature and total population for early time}\label{fig:9d}
        \hspace*{5mm}
    \end{subfigure} 
    \caption{\textbf{Example Bottleneck Planet}: Snapshots of a single experiment showing the geochemical network in black solid lines, and the biochemical network in green dashed lines. Red circles represent source chemicals. $t_{seed}$ is the time the planet was seeded with life. Plot c) shows temperature (red) and total population (black) against time. C = 0.4. The thick end of the geochemical links indicates positive direction.}
\label{fig:9}
\end{figure}

Figure \ref{fig:9} shows an experiment for the same planet as in Figure \ref{fig:8}. This time life failed to survive the bottleneck. We see a very similar pattern as in Figure \ref{fig:8} however importantly the microbes in this experiment fail to evolve a metabolism to consume the chemical species 8. The system can survive a while, compensating for the buildup of chemical 8 by depleting other atmospheric chemicals, however without full control over the atmospheric chemical make-up, the microbes are unable to prevent $T_{planet}$ from rising, and life goes extinct.

Bottleneck planets share the characteristic of having two places where chemicals can accumulate. They otherwise feature many purely geochemical recycling loops. The bottleneck effect occurs early on when life must gain control over the two chemical species with accumulating chemicals; if successful, the recycling loops in the geochemistry prevent the system from fluctuating as wildly as seen in Critical planets. After seeding, Bottleneck planets typically experience a population burst followed by a rapid population decline, before stabilising to a relatively constant total population. The temperature fluctuates the most during this early seeding period. Bottleneck planets can experience population spikes at later times but they are not as severe as seen for Critical planets (Figure \ref{fig:7}) and do not carry the same risk of extinction. Bottleneck planets must also have a geochemistry that allows the temperature to rise fast enough following the cooling caused by the early population burst to prevent inevitable extinction, as seen on Doomed planets (Figure \ref{fig:6}).

\subsubsection{Abiding planets}

Abiding planets are always successfully colonised by life which then goes on to enjoy long-term habitability for every experiment. Abiding planets provide many purely geochemical recycling loops making the system less prone to perturbation than Critical planets for example, however microbe intervention is still required for continued habitability. One simplification of ExoGaia is that geochemical reactions are temperature independent which prevents abiotic temperature feedback loops. Without the influence of life, the vast majority of Abiding planets will quickly reach inhospitable temperatures during their atmospheric evolution. Therefore, while the presence of many geochemical recycling loop can greatly improve the long-term habitability chances of an inhabited planet, on an uninhabited planet there is no temperature feedback process, and thus nothing to prevent temperatures rising to inhospitable abiotic temperatures.

\begin{figure}[htbp!]
\centering
    \begin{subfigure}{0.49\columnwidth}
        \centering
        \makebox[0.5\textwidth][c]{\includegraphics[width=1.0\textwidth]{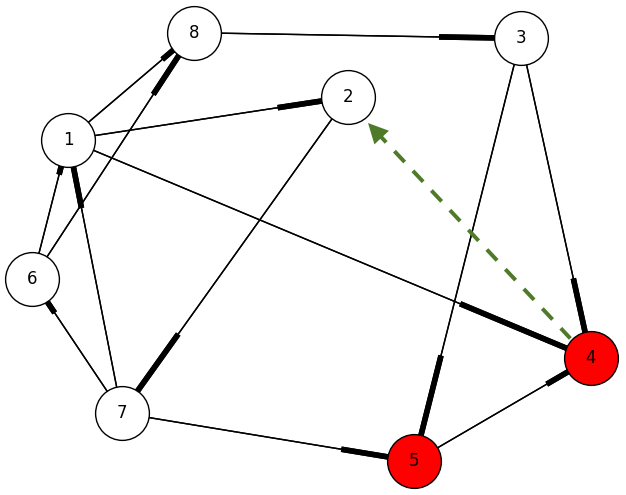}}
       \caption{$t_{seed}$} \label{fig:10a}
        \hspace*{5mm}
    \end{subfigure} 
        \begin{subfigure}{0.49\columnwidth}
        \centering
        \makebox[0.5\textwidth][c]{\includegraphics[width=1.0\textwidth]{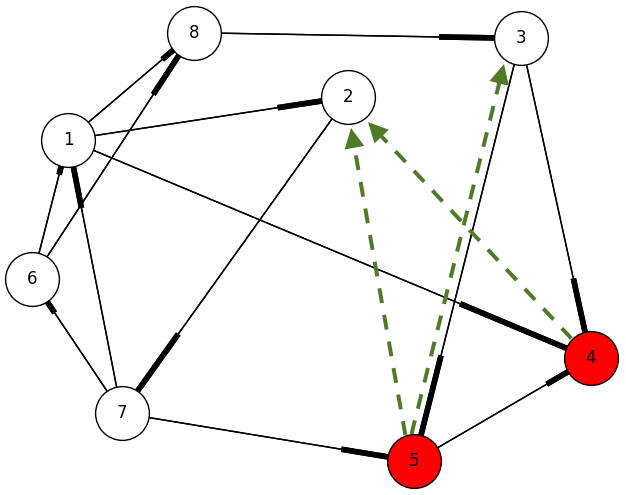}}
        \caption{$t_{seed} + 100$}\label{fig:10b}
        \hspace*{5mm}
    \end{subfigure} 
            \begin{subfigure}{0.49\columnwidth}
        \centering
        \makebox[0.5\textwidth][c]{\includegraphics[width=1.0\textwidth]{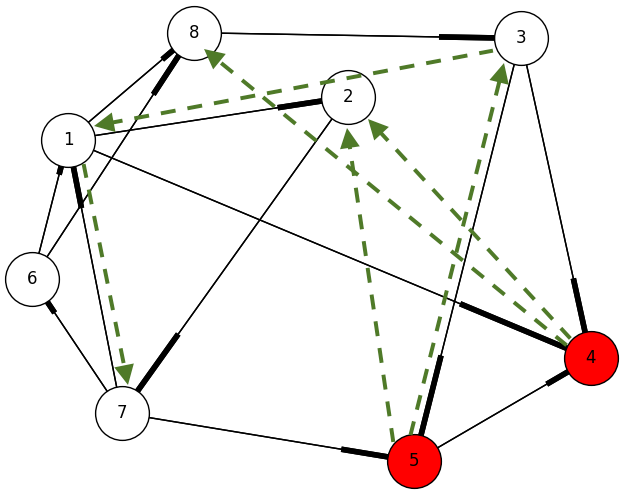}}
        \caption{$t_{seed}+500$}\label{fig:10c}
        \hspace*{5mm}
    \end{subfigure} 
            \begin{subfigure}{0.49\columnwidth}
        \centering
        \makebox[0.5\textwidth][c]{\includegraphics[width=1.0\textwidth]{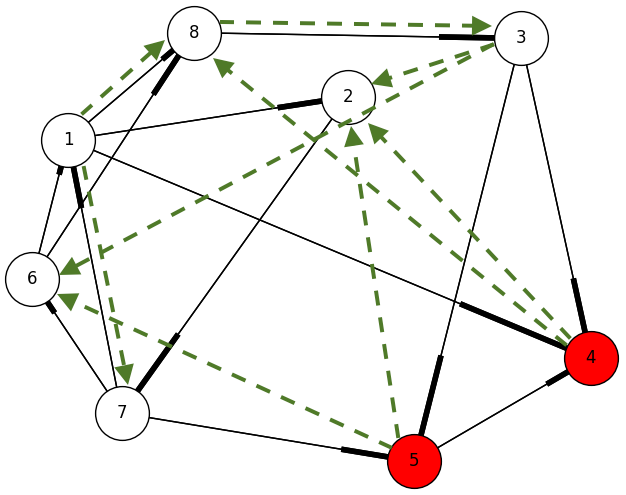}}
        \caption{$t_{seed}+1000$}\label{fig:10d}
        \hspace*{5mm}
    \end{subfigure} 
                \begin{subfigure}{0.49\columnwidth}
        \centering
        \makebox[0.5\textwidth][c]{\includegraphics[width=1.0\textwidth]{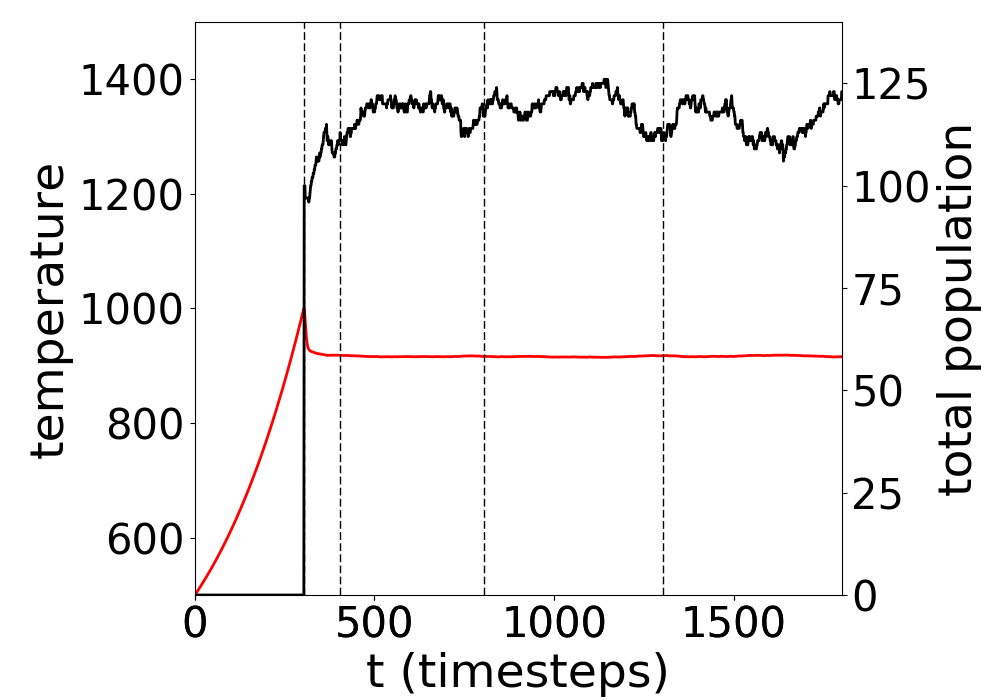}}
        \caption{Temperature and total population for early time \newline}\label{fig:10e}
        \hspace*{5mm}
    \end{subfigure} 
                \begin{subfigure}{0.49\columnwidth}
        \centering
        \makebox[0.5\textwidth][c]{\includegraphics[width=1.0\textwidth]{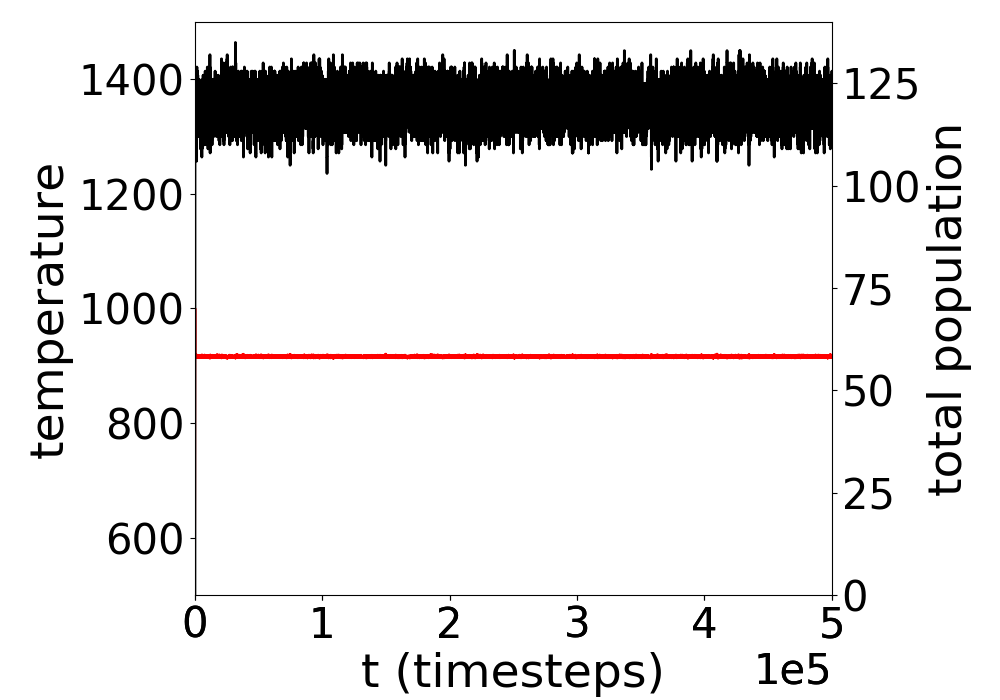}}
        \caption{Temperature and total population for full \newline experiment}\label{fig:10f}
        \hspace*{5mm}
    \end{subfigure} 
\caption{\textbf{Example Abiding Planet}: Snapshots of a single experiment showing the geochemical network in black solid lines, and the biochemical network in green dashed lines. Red circles represent source chemicals. $t_{seed}$ is the time the planet was seeded with life. Plots e) and f) show temperature (red) and total population (black) against time. C = 0.4. The thick end of the geochemical links indicates positive direction.}
\label{fig:10}
\end{figure}

Figure \ref{fig:10} shows snapshots of the biochemistry on an example Abiding planet. The geochemical network of the planet does not provide recycling loops for chemical species 4, but otherwise the geochemistry is well connected with recycling loops present for all possible microbe metabolisms barring those that excrete chemical 4. Figure \ref{fig:10a} shows the first species seeded on the planet with metabolism $4 \rightarrow 2$. As time progresses, the biochemistry infiltrates more and more of the geochemical network. Figure \ref{fig:10e} does not show the population explosion and fall back seen for the Bottleneck planet; instead the population rises and reaches a steady value and stays there. Figure \ref{fig:10f} shows very little fluctuation in the total population or temperature over time.

Abiding planets all share the characteristics of having abundant, purely geochemical, recycling loops. For nearly all microbe metabolisms there are geochemical loops recycling the waste back to food. Abiding planets also typically either have the chemicals well spread between chemical species, or have only a single chemical species that accumulates at high levels. These properties combined make it very easy for life to gain control of its host planet's atmosphere and retain that control. With many geochemical recycling loops that are not subject to fluctuation as biochemical links are, the system is highly stable and thus life is able to successfully colonise and enjoy long-term habitability on Abiding planets.

\subsection{Planet Class Frequency by Connectivity}

\begin{figure}[htbp!]
\centering
    \begin{subfigure}{0.8\columnwidth}
        \centering
        \makebox[0.9\textwidth][c]{\includegraphics[width=1.0\textwidth]{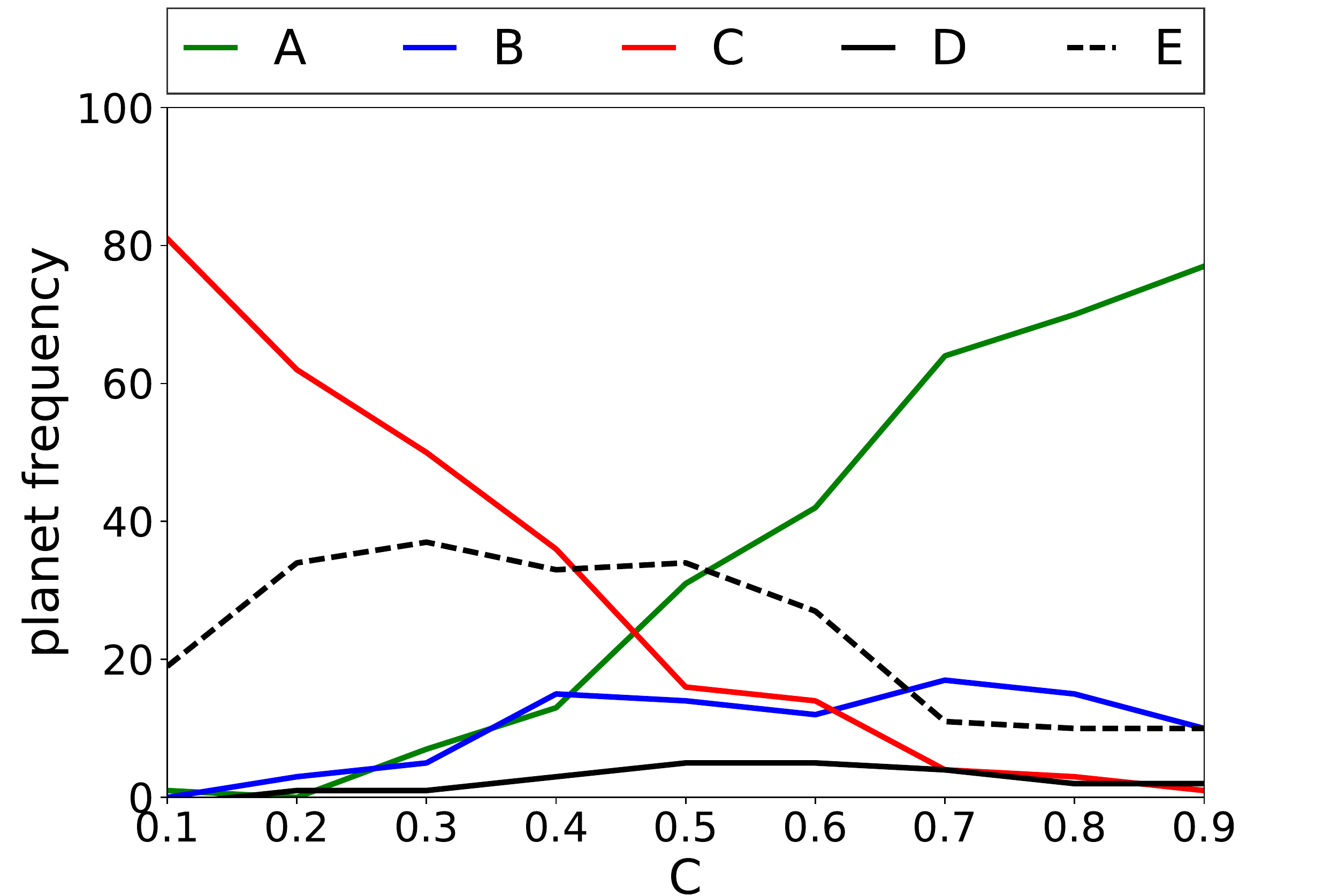}}
        \hspace*{5mm}
    \end{subfigure} 
\caption{The frequency of \textbf{A}biding, \textbf{B}ottleneck, \textbf{C}ritical, \textbf{D}oomed, and \textbf{E}xtreme planets against connectivity for chemical set A}
\label{fig:11}
\end{figure}

Figure \ref{fig:11} shows the frequency of each class of planet against connectivity, $C$. We see a general trend of Abiding planets dominating at high connectivity, Bottleneck planets present mainly at mid and high connectivity, and Critical planets dominating for low connectivity. The number of Extreme planets increases for mid connectivity and then decreases again. Doomed planets make up a small fraction of the planets for all $C$.

As an abundance of geochemical recycling loops, coupled with biotic temperature feedback loops, leads to higher rates of long-term habitability, it is clear why planets with higher $C$ are more likely to be Abiding planets. With more geochemical links there is a greater chance of geochemical recycling loops. Decreasing $C$ means fewer geochemical links, therefore Bottleneck and Critical planets become more likely with Critical planets dominating for very low $C$. For low $C$, biology will have to create more recycling loops itself to successfully regulate the planet's atmosphere, making the system more prone to large scale fluctuations that carry a risk of extinction.

As the source chemicals on average insulate, with few geochemical links most planets for low $C$ will be hot enough for successful colonisation, leading to few Extreme planets. As $C$ increases, the probability of insulating chemicals being converted to reflective chemicals increases and thus so does the frequency of Extreme planets. Increasing $C$ further, the chemicals will become more evenly spread between all chemical species in the chemical set. The average abiotic effect of all the chemical species in chemical set A is insulating, and so the frequency of Extreme planets falls. The exact shape of the planet frequency against $C$ curves in Figure \ref{fig:11} are an artefact of the chemical set used. However, as they are the result of an abstract model, they cannot correspond to any real world data, and we have only one data point to compare to in any case - Earth. The important feature of ExoGaia is that these planet classes emerge, not the relative frequencies of each. The supplementary data for this paper explores alternative chemical sets to demonstrate that chemical set A is not a special case.

\subsection{Planets with habitable $T_{abiotic}$}\label{habitable_t_abiotic}

A small number of modelled planets have habitable $T_{abiotic}$ values. We can compare how the habitability of these planets compares to those planets with $T_{abiotic}$ values that are too hot for life - `hot' planets.
Hot planets will have passed through $T_{pref}$ in their past allowing for seeding; in order to survive, life will have to take control of its planet's atmosphere to maintain habitable conditions and prevent the temperature from rising to the inhabitable $T_{abiotic}$. Table \ref{table:3} lists the number of planets that have a habitable $T_{abiotic}$ for each connectivity, and compares this number to the number of `hot' planets. Table \ref{table:3} shows that the habitable $T_{abiotic}$ planets only make up a small percent of the potentially habitable planets.

\begin{table}
\centering
\caption{The number of planets with habitable $T_{abiotic}$ and number of hot planets for all $C$}\label{table:3}
 \begin{tabular}{ccc}
    \hline
    C & N$^{o}$ habitable $T_{abiotic}$ & N$^{o}$ hot planets \\ \hline
    0.1 & 1 & 80 \\
    0.2 & 4 & 62 \\
    0.3 & 3 & 59 \\ 
    0.4 & 6 & 61 \\
    0.5 & 7 & 59 \\
    0.6 & 13 & 60 \\
    0.7 & 5 & 84 \\
    0.8 & 3 & 87 \\
    0.9 & 2 & 88  \\ \hline
     \end{tabular}
\end{table}

Comparing to Figure \ref{fig:11} we see that the frequency of Critical, Bottleneck, and Abiding planets is far higher than the number of planets with habitable $T_{abiotic}$ values for each $C$, demonstrating that microbes are frequently successful in colonising planets during a short time period of habitability and then acting to prevent temperatures from rising to inhospitable $T_{abiotic}$ values. For mid and high connectivities where we see large numbers of Bottleneck and Abiding planets we see that life can not only colonise planets with inhospitable $T_{abiotic}$, but can maintain long-term habitability. This demonstrates that the microbes can be very successful in regulating their planet's atmosphere. 

Of the planets with habitable $T_{abiotic}$ values listed in Table \ref{table:3}, only one, for $C = 0.4$ was an Abiding planet. None were Bottleneck planets; the majority were found to be Critical and Doomed planets. This shows that planets where $T_{abiotic}$ is habitable are in fact not generally planets that support life long-term. The reason for this is as outlined in Section \ref{subsubsec:Dead} for the example Doomed planet - life must be able to remove chemicals from the atmosphere to metabolise and survive, and doing so must not push the planet beyond the bounds of habitability. If a planet has a $T_{abiotic} \approx T_{pref}$ then removing chemicals is highly likely to decrease habitability, rather than maintain it (as is the case on many `hot' planets) thus making such planets, somewhat counterintuitively, mostly poor candidates for long-term habitability.

\section{Discussion}

The ExoGaia model demonstrates planetary temperature regulation, performed by a simple biosphere. There are two extinction mechanisms in ExoGaia - planetary over cooling caused by microbe activity, or over heating due to abiotic processes following the loss of biotic atmospheric control. Under favourable conditions, life on an ExoGaia planet can enjoy long-term habitability and can prevent temperatures from rising to inhospitable levels as would happen on a planet devoid of life. For colonisation success, microbes require the host planet's temperature to reach a preferred temperature, $T_{pref}$, during its atmospheric evolution, and require a geochemical network that allows temperatures to recover fast enough after microbe induced cooling to avoid microbe extinction. For long-term habitability, microbes require a planet with a geochemical network that provides many chemical recycling loops. By seeding planets at $T_{pref}$ we have investigated the microbes' ability to maintain the planetary temperature within habitable bounds. The ExoGaia model demonstrates that apparently complex global phenomena such as regulation can arise from the simple interaction of the small parts making up a system. Five distinct planet classes emerge from the ExoGaia model:

\begin{itemize}
\item \textbf{Extreme} - Planets that never reach habitable temperatures
\item \textbf{Doomed} - Planets that reach habitable temperatures but are unable to support life.
\item \textbf{Critical} - Planets that have a higher colonisation success than long-term habitability success. 
\item \textbf{Bottleneck} - Planets that if successfully colonised enjoy long-term habitability.
\item \textbf{Abiding} - Planets that are always successfully colonised and always have long-term habitability. 
\end{itemize}

We can consider what these results might imply for real planets. Our model predicts that more geologically active planets may be more suitable hosts for life. More geochemical processes provide more potential chemical recycling networks for life to exploit and our model biospheres are more adept at dampening or accelerating pre-exiting geochemical reactions than at forming stable stand alone chemical links. There are clear real world examples however where biological processes are dominant, i.e. the concentration of oxygen in our atmosphere, highlighting the limits of our model for application to the real world.

Which model planet class might Earth belong to? Clearly we do not live on an Doomed or an Extreme planet. We also do not see frequent rapid very large-scale changes in the total population of the biosphere of Earth, perhaps making it unlikely that Earth is a Critical planet. The mass extinctions during the Phanerozoic \cite{Raup1982}, were not the regular large-scale stochastic fluctuations typical of our model Critical planets, but rather more akin to regime shifts between periods of quasi-stability. Many of the suspected triggers for these mass extinctions are abiotic phenomena excluded from the ExoGaia model, such as meteor impacts, volcanic events, and changing sea levels \cite{White2005}. These extinctions were also mainly - but not exclusively - of macroscopic organisms, which are a tiny percentage of the biodiversity on Earth even today; from the point of view of microbes, making up the majority of Earth's biomass, these events would probably not be classed as mass extinctions \cite{Nee2004}. If Venus and Earth are alternate states of the same system \cite{Lenardic2016} perhaps we are on the lucky side of a Gaian bottleneck? We know that certain biological innovations, e.g. the evolution of oxygenic photosynthesis \cite{Hoffman2013}, and later on the evolution of land plants \cite{Lenton:2012aa}, likely triggered ice ages, the former as oxidation of the atmosphere mediated collapse of a $CH_{4}$ greenhouse effect, and the latter as land plants increased weathering thus increasing the rate of $CO_{2}$ removal from the atmosphere. This is perhaps similar to the cooling some Bottleneck planets experience when life is first established. Models of the habitable zone under purely abiotic control, e.g. carbonate-silicate weathering, predict that Earth would be habitable without life (e.g. \cite{Kopparapu2013}). When examining planets with habitable $T_{abiotic}$ values in Section \ref{habitable_t_abiotic} we saw Critical and one Abiding planet represented. This could suggest that Earth might be an Abiding planet. 

Venus' current inhospitable state could indicate it being on the `losing' side of a Gaian bottleneck as previously speculated, or could indicate a break down of regulation being performed by a hypothetical Venusian biosphere, making Venus a Critical planet. There is no data on how a life-environment coupled Venus system would behave over long time periods, preventing the sort of analysis possible for Earth. If the runaway greenhouse that occurred on Venus was unavoidable, as many models suggest (e.g. \cite{Kopparapu2013}), then Venus would perhaps most closely correspond to a Doomed planet due to the evidence that it once hosted liquid water (\cite{Donahue1982, Jones2003}) and thus may have once been potentially habitable. Changes in solar luminosity were not considered within the ExoGaia model, and so planets that might have hosted a biosphere, and then lost habitability through unavoidable external factors, do not fit well into the model planet classification system.

We can also consider Mars as observational evidence points to it once having had large bodies of liquid water, e.g. \cite{Milton:1973aa}. It is not known what the early environment of Mars was like, whether it was warm and wet \cite{Craddock:2002aa}, or cold with volcanism and impacts causing transient warm conditions \cite{Wordsworth:2013ab}. If the latter, potential habitats for Martian life might have been heterogenous throughout time and space, possibly preventing any early life from spreading across the planet \cite{Cockell:2012aa}. If this were the case, Mars might most closely correspond to a Doomed planet - a window of habitability existed, however life was unable to flourish. If Mars did at one point host a substantial biosphere, it has clearly lost it. Mars once had a far thicker atmosphere which it has since lost \cite{Pepin:1994aa}, causing the dry cold conditions on Mars today. Atmospheric loss was not taken into account in the ExoGaia model, however this could perhaps be very loosely compared to an uncontrolled build-up of a cooling chemical on a model planet that a biosphere might mitigate for a while, potentially making Mars a Critical planet. However, Critical planets are theoretically habitable indefinitely, while any planet undergoing significant atmospheric loss will experience drastic changes in its surface environment, making this comparison far from ideal. There is ongoing speculation that life might yet be found on Mars in sparse pockets \cite{Wilkinson2006}. ExoGaia is mainly concerned with large-scale planetary regulation, and therefore small refuges of life with little to no impact on global parameters are predicted to impact model results only if conditions improved to allow this life another chance of becoming globally established (see Appendix B for experiments along this theme).
 
With a highly simplified and abstract model like ExoGaia, no strong predictions can be made for individual planets, and comparisons between real planets and model planet classifications highlight the many limitations of the model. More complex future versions of ExoGaia could begin to address some of the questions raised by considering specific planets within the ExoGaia framework and future space missions to Venus and Mars might provide more data to compare with model planet classifications. It is difficult to determine which class a planet might fall into based on a single time point; the planet classes in ExoGaia are best identified by looking at the whole planet history. Therefore, any methods that can provide long timescale observations of planets would provide the best data for comparison with model predictions.

The ExoGaia model adds to the narrative that for a planet to remain habitable, it must be inhabited \cite{Lenardic2016}. It suggests that geologically active planets still early in their atmospheric evolution would be the most suitable candidates for colonisation by life and agrees with the idea that when searching for inhabited exoplanets we should look for planets with atmospheres in disequilibrium \cite{Lovelock1965}. Our model suggests that many planets that have had life will have lost it, however that some, with the right geological conditions, can enjoy long-term uninterrupted habitability. Currently with only one data point - the Earth, we cannot draw any conclusions. As more exoplanets are found, their macro properties determined, and their atmospheres analysed, we will have more data available to compare with model predictions.

Further work should explore how the ExoGaia model behaviour is impacted by adding temperature dependant abiotic processes, and the effects of changes in solar luminosity or other abiotic perturbations. Our model microbes could also be made more complex, as microbes can be found in almost any part of our globe, from the Mars-like conditions of the Antarctic dry valleys \cite{Siebert1996} to hydrothermal vents at around 122$^{o}$C \cite{Clarke2014}, a fact not reflected in our model where microbes have a universal temperature preference. Adding spatial structure to models has been shown to be very important in work in theoretical ecology over recent decades \cite{Nee2007} and therefore is an obvious next step in developing this model. Introducing spatial heterogeneity into the model would also allow life to seek refuges during periods of extreme climate change, similar to how life is thought to have survived in small oases during the snowball earth events, or speculated to possibly persist on Mars today. The change in model dynamics in response to adding spatial structure would be an important next step in improving the applicability of the ExoGaia to real planets.

\section*{Acknowledgements}

We thank the Gaia Charity and the University of Exeter for their support of this work.

\bibliographystyle{apalike}
\bibliography{ExoGaia_arXiv} 

\newpage
\appendix

\section{ExoGaia Model Description}
\label{appendixA}

Code made available upon reasonable request to corresponding author.

The ExoGaia model uses agent based dynamics to describe a biosphere consisting of simple microbes interacting with a host planet via consumption and excretion of atmospheric chemicals. These chemicals determine the surface temperature of the planet. In this appendix each part of the model will be described in detail and then the experiment method will be presented at the end.

\subsection{Microbes}

The  microbes consume chemicals as food and excrete chemicals as waste products. A particular microbe's food and waste product are encoded in the genome of each microbe species. All microbes share the same ideal temperature (i.e.  the temperature which results in the maximum growth rate). Microbes grow by consuming chemicals and  converting them to biomass. They reproduce  asexually by splitting once their biomass reaches a threshold. Biomass is decreased by a fixed amount per timestep to represent the cost of staying alive. Microbes die if their  biomass drops below a fixed threshold, which can happen due to food limitation or temperature limitation leaving the microbes unable to consume the chemicals present.

In the code we do not record microbes of the same species individually as doing so would slow the simulations considerably. Instead we group microbes of the same species together and  record the species' total biomass. Thus each species can be thought of as a list $M$:

\begin{equation}
\label{eq:A1}
M = (g, N, B, F, W, T_{pref})
\end{equation}

where $g$ is the species' genome (represented as a decimal number), $N$ is the population of the species, $B$ is the total biomass of the species, $F$ is total number of consumed food chemicals not yet converted into biomass, $W$ is the total number of waste chemicals not yet excreted by members of the species, and $T_{pref}$ is the temperature that maximises the growth rate for species $M$. All species share the same $T_{pref}$.

\subsubsection{Genotype}

The genotype of a microbe is recorded  as the decimal representation of an 8 bit binary  string, and this is used to group microbes into species. Microbes that share the same genome are of the same species. We create tables for microbe chemical consumption and excretion rules, and this genome is used as the reference to look up the particular metabolism for a microbe. These tables are generated in the following way: for each possible genome, a chemical species is selected at random to be the food source for microbes of that genome. Another chemical species is then selected at random to be the waste for microbes of that genome. The food source and waste of a microbe must not be the same, so if the waste chemical species selected is the same as the food, another chemical species is chosen at random until these are not the same. All microbes consume only one type of chemical and excrete only one type of chemical. The index of a microbe's metabolism in the table is the decimal value of the microbe's genome.  With an 8 bit long binary genome there are 256 possible species (as each gene in a genome can have the value 0 or 1).

Table \ref{table:A1} shows an example look up table. To use Table \ref{table:A1}, for a microbe with genome 000000010, we convert to its decimal value, 2, and find that this microbe has metabolism $1 \rightarrow 2$ i.e. it consumes chemical species 1 and excretes chemical species 2.

\begin{table}
\centering
\caption{Example microbe metabolism look up table}
\label{table:A1}
 \begin{tabular}{ccc}
    \hline
    Index & Food Chemical & Waste Chemical \\ \hline
    0 & 2 & 6 \\
    1 & 4 & 1 \\
    2 & 1 & 2 \\ \hline
     \end{tabular}
\end{table}

\subsubsection{Chemical Consumption}

When a microbe is selected to consume, it will attempt to eat $K_{j}$ units of its chemical food source  (the value of $K_{j}$ depends on how closely the planetary temperature matches the microbes' preferred temperature, and the microbes' sensitivity to the environment), and will be successful if the chemicals are available. 

For simplicity we limit our microbes to single chemical metabolisms, meaning that a microbe  only consumes one type of chemical,  and only excretes one type of chemical, with the limitation that no microbe may consume what it excretes.

\subsubsection{Metabolism}

The microbes convert their food into biomass in an inefficient process that produces waste product. The efficiency of this conversion is given by $\theta$, and the amount of biomass produced is given by: 

\begin{equation}
B_{j} = \theta F_{j}
\end{equation}

where $B_{j}$ is the number of biomass units produced and $F_{j}$ is the number of food units currently `contained' with a microbe $j$. The waste excreted in this process is given by: 

\begin{equation}
W_{j} = (1- \theta) F_{j}
\end{equation}

where $W_{j}$ is the number of waste units produced, which are released into the environment after the biomass has been created, in the form of the chemicals determined by microbe $j$'s specific metabolism (e.g. see the look up table example in the previous section).

\subsubsection{Effect of temperature on metabolic rate}

The  state of the abiotic  environment affects the rate at which  microbes  can  consume  chemicals which  in  turn affects the rate  of biomass  production and  thus the growth of the microbes. A microbe will attempt to consume an amount of chemicals $K_{j}$ each timestep  with  the demand  being met  depending  on chemical  availability. $K_{j}$ is calculated for each microbe j as a function of the difference between the microbes' ideal temperature and the current planetary temperature. This  function is has a Gaussian form and falls away smoothly from its maximum as the distance between the optimum and the current environment increases. This is a widely used assumption when modelling an organism's response to the temperate of its environment. Mathematically we write this as:

\begin{equation}
K_{j} = \psi_{j}K^{max}
\end{equation}

\begin{equation}
\psi_{j} = e^{-(\tau p_{j})^{2}}
\end{equation}

\begin{equation}
p_{j} = \sqrt{(T_{planet} - T_{pref})^{2}}
\end{equation}

where $K^{max}$ is a constant determining the maximum  rate  of consumption for any microbe, $\psi_{j}$ is a microbe specific measure of the microbe's satisfaction with the current abiotic  environment, $\tau$ is a universal  constant parameter that determines how sensitive the microbes are to their environment ($\tau = 0$ means the microbes are not affected by the abiotic environment at all, and a higher $\tau$ means the microbes  become  more  sensitive  to the abiotic  conditions). The effects of changing this $\tau$ parameter has on system dynamics has been explored in the Flask model (e.g. \cite{Nicholson2017}) on which this model is heavily based on. $p_{j}$ is a measure  of the (positive) distance between  the current  environmental  temperature, $T_{planet}$, and the microbe's preferred temperature, $T_{pref}$.

\subsubsection{Maintenance Cost}

There is a fixed biomass cost $\lambda$ of staying alive for each microbe. This  reduces  a microbe's  biomass by a constant rate.  This  cost represents the energy costs of maintaining cellular  machinery  and metabolic inefficiency.  This cost is assumed to be lost as unrecoverable heat radiation. This ensures that the chemicals cannot be infinitely recycled and it sets the carrying capacity of the system. This carry capacity is reached  when the total heat  dissipation matches  the energy supplied  in the form of chemicals, i.e. the food the microbes consume. As any heat dissipation of the microbes in the real world due to metabolic inefficiency is many orders of magnitude smaller than the effects of the atmospheric composition on planetary temperatures, we neglect this heat dissipation when calculating planetary temperatures.

\subsubsection{Reproduction and Mutation}

If the microbe is able to consume enough chemicals to reach the reproduction threshold $T_{R}$,  it will reproduce asexually, splitting in half.  Half of the biomass with go to the new microbe and the parent microbe will be left with half its biomass. The new microbe will have the same genome as the parent unless a mutation occurred during  the reproduction. There is a small constant probability of mutation, $P_{mut}$, for each locus. During a reproduction event, the code iterates through the genome of the new microbe and if a mutation occurs at a locus then the gene at that point will be `flipped', turning it to 0 if it were previously 1, or to 1 if it were previously  0. This new mutant genome will then dictate the new microbe's metabolism.

\subsubsection{Death}

If a microbe's biomass falls to a starvation threshold $T_{D}$  the microbe will starve to death. There  is another small probability of death $P_{D}$ that represents death by hazardous mutation or damaging local environmental changes etc. When a microbe dies its biomass is removed from the system, as if the dead microbe, for example, fell to the bottom of the ocean. During a death event, we first check to see if the selected microbe has enough biomass to avoid death by starvation. If the microbe has not starved to death it will be killed with probability $P_{D}$. 

\subsection{Selecting a microbe}

We use agent based dynamics in our model. This means within a timestep, a microbe is chosen randomly for an event and time is effectively frozen while the microbe performs that event. Time is then restarted and another microbe is chosen at random for an event. 

As we record microbes grouped together in a species (Equation \ref{eq:A1}), for any particular species we have the population of the species, the total species biomass, and the total consumed food not yet converted into biomass. To select a single individual of a particular species we therefore need to determine how much biomass and unconverted food this individual has. If a microbe is selected for a reproduction event, we need to know how much biomass it has to know if it has reached the reproduction threshold for example. 

There will be variation between individuals of a species and so we assume a normal distribution of biomass and unconverted food between individuals of a species. The biomass normal distribution is centred around the average amount of biomass $B_{av}$ per microbe (i.e. the total species biomass divided by the species population), with standard deviation of the distribution is $B_{av} \times 0.1$. The normal distribution for the unconverted food is the same but with $F_{av}$, the average amount of unconverted food per microbe, instead. The standard deviation for both distributions is small, resulting in a small level of variation in the population. Therefore most individuals of the same species will have the same biomass and food levels.

Once we have selected a microbe and calculated its biomass and food level, the microbe can then attempt to perform the event it was selected for.

\subsection{Planet setup}

Each planet has a well mixed atmosphere with no spatial element. The atmosphere is characterised by chemicals. There are 8 possible chemicals in ExoGaia, although not all chemicals have to be present in the atmosphere at the same time. The chemicals present in the atmosphere may be consumed by microbes and converted into biomass, and the atmospheric chemical composition determines the temperature of a planet.

\subsubsection{`Temperature' in ExoGaia}
\label{SectionA1.3}

When calculating temperatures in the ExoGaia model we make a simplification of the Stefan-Boltzmann law. Instead of $\beta \propto T^{4}$ , where $\beta$ is the incoming energy to the planet from the `star' and $T$ is planetary temperature, we simplify to $\beta \propto T$. This approximation has been used before in Daisyworld to make determining the underlying regulation mechanisms easier. It has been noted (\cite{Watson1983} \cite{Saunders:1994aa} \cite{De-Gregorio:1992aa} \cite{Weber:2001aa} \cite{Wood:2006aa}) that this simplification does not greatly change the overall behaviour of the Daisyworld model. The Stefan-Boltzmann equation is close to linear at real world habitable temperatures, i.e. near 22$^{o}$C. In ExoGaia, we are only interested in planetary dynamics when there is life on a planet, so while the `temperature' in the ExoGaia model is not constrained, we are only interested in a narrow range of temperatures where life is possible. The temperature behaviour outside this range is not important to the model results. We use an unrealistic $T_{pref}$ for our model microbes to highlight the abstract nature of the model, however as a near linear relationship exists at habitable conditions on Earth, and we are striving to simplify the model abiotic environment as much as possible, we take $\beta \propto T$, where $\beta$ is the energy provided to the planet by the host star per timestep, $T$ is temperature. We then make a further simplification and take the value of $\beta$ to be equal to the value of $T$.

\subsubsection{Chemical Species}

In ExoGaia we have different `chemical species' as an abstract representation of real-world atmospheric gases e.g. $CO_{2}$, $CH_{4}$, or $O_{2}$. These abstract chemical species are not meant to mimic any specific real world chemistry. Each chemical species insulates or reflects by a particular amount. The maximum reflective or insulating property of a chemical species $i$ is represented by $a_{i}$. These $a_{i}$ values are taken from the range [ -1, +1]. A negative $a_{i}$ corresponds to a reflective chemical species, and a positive $a_{i}$ means it is insulating. A positive $a_{i}$ might represent for example the maximum insulating effect of an atmosphere saturated with $CO_{2}$.  The strength of the effect exhibited by any chemical species, $S_{i}$, depends on the number of particles of that chemical in the system, e.g. the abundance of $CH_{4}$ say in the atmosphere:

\begin{equation}
\label{eq:Ax}
S_{i} = a_{i} \text{ tanh}\left(\frac{n_{i}}{D}\right)
\end{equation}

where  $n_{i}$  is the abundance of chemical species $i$, i.e. the number of particles of chemical $i$ present in the atmosphere, and  $D$  is a large number  to make  the effects  of a single `particle' of each chemical species small. This enables large populations to be supported where the individual effect of a single microbe's consumption and excretion of chemicals is small. We use tanh as it is a function that smoothly varies between 0 and 1. The maximum effect any chemical species, $i$, can have is determined by its $a_{i}$ value and by using tanh we can cap the reflective or insulating effects of a chemical species to its $a_{i}$ value. This does not prevent runaway temperature changes in the model, as seen when planetary temperatures rise to above the microbe's ideal temperature.

A subset of chemical species are chosen as `source chemicals'. These are chemical species with an inflow from what  we could think of as the `mantle' of the planet, e.g. $CO_{2}$ from volcanoes.  Each source chemical has a constant inflow rate  $I_{N}$, and there are $N_{S}$ source chemicals. This inflow is kept at a constant rate per timestep for the full experiment. Any chemical species that is not a source chemical does not exist in the atmosphere unless it is produced by a geochemical or biochemical process. 

\subsubsection{Atmospheric properties and planetary temperatures}
\label{MD:atmos}

The state of the atmosphere is given by a vector V:

\begin{equation}
V = (n_{1}, ... , n_{N})
\end{equation}

where $n_{i}$ is the abundance of chemical species $i$, and $N$ is the number of chemical species. As each chemical species in the model has an insulating or a reflective property, the planet atmosphere's insulating or reflective effect will depend on the chemical composition of the atmosphere.

We define $A_{I}$ as the fraction of the planet's current thermal energy retained by the atmosphere via insulation, and $A_{R}$  as the fraction of incoming solar radiation reflected by the atmosphere. The total reflective and insulating  properties of the atmosphere depends on the amount of each type of chemical present. We calculate $A_{R}$, and $A_{I}$ in the following way:

\begin{equation}
A_{R} = \sum_{i \in R}  - a_{i} \text{ tanh}\left(\frac{ n_{i} }{D}\right)
\label{equation:18}
\end{equation}

\begin{equation}
A_{I} = \sum_{i \in I} a_{i}  \text{ tanh}\left(\frac{ n_{i} }{D}\right)
\label{equation:19}
\end{equation}

$R$ is the set reflective chemical species and $I$ is the set of insulating chemical species. $n_{i}$ and $D$ are the same as for Equation \ref{eq:Ax}. $A_{R}$  and $A_{I}$  are constrained to be between 0 and 1, as the maximum amount of thermal energy a planet can retain is the energy it currently has, and the maximum amount of incoming radiation that can be reflected is the amount incoming from the host star, so we also have:

\begin{flalign}
\text{if}\ A_{R} > 1 \rightarrow A_{R} = 1\\
\text{if}\ A_{I} > 1 \rightarrow A_{I} = 1
\end{flalign}

We define  $\beta_{planet}$ as the planetary thermal energy and $\beta_{star}$ as the incoming solar radiation per timestep. We then calculate  $\beta_{update}$, the updated thermal energy of the planet including the insulating effect of the atmosphere in the following way:

\begin{equation}
\label{equation:A13}
\beta_{update} = A_{I} \beta_{planet} + (1 - A_{R})  \beta_{star}
\end{equation}

Using the simplification in Section \ref{SectionA1.3}, the $\beta$ values correspond to temperature values, so that if the thermal energy of a planet was $\beta_{planet}$, then the value of $\beta_{planet}$ will be the same as the value of $T_{planet}$ - the temperature of the planet.

\subsubsection{Chemical inflow and outflow}

There is a constant rate of inflow of source chemicals. Each timestep, $I_{N}$ particles of each source chemical will be added to the system. There is a rate of outflow from every chemical species that is abundance dependant. Each timestep every chemical species will experience an outflow of $n_{i} \times O_{N}$ where $n_{i}$ is the abundance of chemical species $i$, and $O_{N}$ is a constant rate of outflow. Therefore more abundant chemical species will experience a higher rate of outflow than less abundant ones.

\subsubsection{Geochemistry setup}

Each planet has geochemical reactions taking place throughout the experiment. For our geochemistry, we have links between chemical species converting one chemical type to another. The  process is assumed  to be 100$\%$ efficient, so one particle of chemical A would be converted to one particle of chemical B. Links between chemical species can only flow in one direction, so if we have a process converting chemical $A \rightarrow B$, we cannot  then have another geochemical process converting $B \rightarrow A$.  Other routes are allowed though, i.e.  $B \rightarrow C \rightarrow A$ for example. This simplification makes it simpler to track chemicals as they move through the system. Real world systems have chemical reactions that can be reversed, however we could also consider this simplification to be the net movement of chemicals once each direction of the reaction has been taken into account. If $A \rightarrow B$ and $B \rightarrow A$, we can still describe the overall movement of chemicals between A and B with a link of either $A \rightarrow B$  or $B \rightarrow A$.

Geochemical reactions take place at a rate that depends on the abundance of the reactant chemical species. Each geochemical link is randomly assigned a value taken from the uniform range $[0,1)$ which we call the `link strength'. This number determines what percentage of the reactant chemical species is converted to the product chemical species due to the geological process, per timestep. E.g. if we have a geological link: $A \rightarrow B$, with strength 0.2, this means every timestep 20$\%$ of chemicals type A are converted into chemicals of type B. 

If we have a matrix  $G$ that represents a planet's geochemical reactions, $G_{ij}$ would be the flow from chemical species $i$, to chemical species $j$ due to a geochemical reaction. If $G_{ij} > 0$, then $G_{ji}  = 0$ as we don't allow for links flowing between the same two chemical species in opposites directions. For a particular connectivity, say $C = 0.1$, each chemical species has $10\%$ chance of being connected to another. We then determine the strength of the connection, i.e. how fast the process is that converts A to B. We set up our geochemical processes in the following way.

To populate the geochemical reaction matrix $G$, we consider each pair of chemical species in turn. The connectivity $C$ tells us the probability that these two chemical species will be connected by a geological reaction, or link. We generate a random number $r_{1}$ taken from the uniform range $[0,1)$, and if $r_{1} < C$ then our chemical species are connected by a link. If $r_{1} \geq C$ the two chemical species are not connected.

If the chemical species are connected we then generate another random number, $r_{2}$, also from the uniform range $[0,1)$ to determine which direction the link flows in, e.g. $A \rightarrow B$ or $B \rightarrow A$ with each direction having equal probability. 

Once the direction of the link is determined, the strength of the link is then found by generating a third random number, $r_{3}$, (from the uniform range $[0,1)$ ) and the link strength $L_{s} = r_{3}$. We repeat this process for each pair of chemical species.

Thus we end up with a matrix  G of the following form:

\begin{equation}
G = 
\begin{bmatrix}
0 & 0 & a_{2,0} & 0 & 0 \\
a_{0,1} & 0 & 0 & a_{3,1} & 0 \\
0 & a_{1,2} & 0 & a_{3,2} & 0 \\
0 & 0 & 0 & 0 & a_{4,3} \\
0 & a_{1,4} & 0 & 0 & 0 \\
\end{bmatrix}
\end{equation}

G contains all the geological processes happening on a planet, with their strength and direction. All the $G_{ii}$ indices are  0, and where  $G_{ij} > 0$ it is always  true that $G_{ji}  = 0$.  If $G_{1,2} = 0.7$ for example, it means  that every timestep 70$\%$ of chemical species 1, will be converted into chemical species 2.

Each timestep we can therefore loop though $G$ to determine where chemicals are moving due to geochemical processes. For a non zero $G_{ij}$ value, chemical species $i$ will be depleted by $n_{i}G_{ij}$ and chemical species $j$ will be incremented by the same amount due to the geological process. We do this for each geochemical process and add up the total amount of chemicals added to or removed from each chemical species for each timestep.

\subsection{Seeding a planet}

A planet is seeded with microbial life when the temperature of the planet $T_{planet}$ equals the microbes' preferred temperature $T_{pref}$. Because of the way temperature is determined in the model, planet temperatures might never exactly match $T_{pref}$, so to ensure that seeding still occurs we determine a suitable `seeding window' $S_{w}$ - a small temperature range close to $T_{pref}$. Seeding can occur when planet matches any temperature in $S_{w}$ but seeding only occurs once. These $T_{pref}$ and $T_{planet}$ temperatures correspond to thermal energies, and using the simplification in Section \ref{SectionA1.3} we can take the values of $T_{pref}$ and $T_{planet}$ to be same as the values of the corresponding thermal energies $\beta_{pref}$ and $\beta_{planet}$.

For the case where $\beta_{star} < \beta_{pref}$ we require an insulating atmosphere for habitability. Therefore we determine our seed window, $S_{w}$, as the range $[T_{pref}, T_{pref} +50]$. As we know for $\beta_{star} < \beta_{pref}$ we must have an insulating atmosphere for potential habitability, the $S_{w}$ range goes higher than the ideal temperatures, so that if temperatures never exactly match $T_{pref}$ as the temperatures continue to rise, seeding still takes place, and life will still be seeded at a hospitable temperature.

For the case where $\beta_{star} > \beta_{pref}$, we need a cooling atmosphere for habitability therefore we set $S_{w} = [T_{pref} -50, T_{pref}]$. The logic is the same, however here, as the atmosphere on a potentially habitable planet in this setup will be a cooling atmosphere, the temperature will be falling when it passes through $T_{pref}$ so we allow for slightly cooler temperatures in case $\beta_{planet} = \beta_{pref}$ in the simulation never takes place exactly.

When $\beta_{star} = \beta_{pref}$, we have an extra requirement that is automatically fulfilled in the previous two scenarios. When we seed with life, we require there to be food for the life to consume. When $\beta_{star}$ is far from $\beta_{pref}$, we know that when the planet's temperature becomes habitable, it is because an atmosphere has built up. When $\beta_{star} = \beta_{pref}$ , seeding could occur when no food was present. To deal with this we add an extra requirement for seeding when $\beta_{star} = \beta_{pref}$. $S_{w}$ is now in the range $[T_{pref} -50, T_{pref} + 50]$ as a potentially habitable planet could have either a cooling or insulating atmosphere, and now we require that at least one chemical species in the system must have an abundance greater than 1000. This means that although conditions will start with $T_{planet} = T_{pref}$ seeding is delayed until there are some atmospheric chemicals present for the microbes to consume. These chemicals will likely alter planetary temperatures and thus degrade the environment, however provided an abundant food becomes available before the seed window is missed, seeding will take place.

For all scenarios, when seeding a planet, we seed with one species and we seed with $M_{N}$ individuals of that species. We choose the seed species at random, however we ensure that the species chosen has an abundant food source available to it. If species $A$ consumes chemical $C_{A}$, if there are greater than 1000 units of chemical $C_{A}$ in the atmosphere at the time of seeding, species $A$ is a suitable seed species. If there are fewer than 1000 units of $C_{A}$ present in the atmosphere at the time of seeding then species $A$ is not a suitable seed species and another species is chosen at random until a suitable species, that consumes a presently abundant chemical, is found. This makes biological sense as a species will not evolve to consume a nonexistent food source.

If a seed window has not been passed after $5 \times 10^{4}$ timesteps then an seeding attempt is made once, and the model then continues as usual for $50 \times 10^{4}$ timesteps.

It seems sensible that life on Earth will have emerged to initially consume something plentiful which is why we take this approach in our model. If life initially emerged to consume something not plentiful then extinction will have quickly followed. As life did indeed emerge on Earth, either it initially had a stable food source, or it emerged many times and went extinct until a life-form that consumed a food source abundantly present emerged and avoided extinction via starvation.  

\subsection{Model Timesteps}

We use agent based dynamics to run the simulation and a timestep is broken down into `iterations'. The number of iterations per timestep depends on the the number of microbes alive in the system at  the start of the timestep. In reality, microbes eat food, create biomass, excrete waste, reproduce, and die all in parallel with one another. The model steps performed within a timestep in the ExoGaia model would also ideally all be computed in parallel but computational limitations prevent this, and so for agent based dynamics we effectively freeze the system while a selected microbe performs an action. Therefore timesteps are broken down into iterations. An iteration consists of the following steps: 

\begin{itemize}
\item A microbe is randomly selected for a chemical consumption event
\item A microbe is randomly selected for a biomass creation event
\item A microbe is randomly selected for a reproduction event
\item A microbe is randomly selected for a death event
\end{itemize}

These events are repeated $N_{M}$ times each where $N_{M}$ is the microbe population at the start of a timestep. A microbe selected for an event will not necessarily perform that event. For example, the microbe might not have enough biomass to reproduce, or the temperature might be too hot or cold for a microbe to consume chemicals. Being selected for an event means that a microbe will perform that event only if conditions allow, and depends on the probability of the event successfully occurring.

We also break down the inflow and outflow of chemicals and to prevent sudden changes at the the start of each timestep. If we simply added and deducted the chemical flow amounts at the start of each timestep, microbes selected at the beginning of a timestep could see a very different world to those selected at the end of a timestep and large sudden changes could occur between timesteps. Although these effects would largely average out due to the random selection of microbes during each timestep, a single large influx per timestep could be thought of as a periodic perturbation on the system which could affect the results seen. To counter this, we calculate the total inflow (from external sources if a source chemical, and from inflows due to geological processes) and outflow of each chemical species at the start of each timestep and divide this by the number of iterations in the timestep, i.e. the microbe population at the start of the timestep, $N_{M}$:

\begin{equation}
N_{i}^{change} = I_{N} - O_{N} N_{i}^{ab} / N_{M}
\end{equation}

$I_{N}$ is the number of units  of chemical  inflow per timestep,  $O_{N}$  is the percentage  outflow, and  $N_{i}^{change}$ is amount we increment chemical species $i$'s abundance each iteration. This results in the same quantity of chemicals  being added  / removed  from the system  as if there was just one update at the start of the timestep, but it results in a much smoother transition and  means  that microbes selected at  the start and end of a timestep will see much  more similar worlds. Of the life events of a microbe, only chemical consumption depends on the external environment which means only one event within an iteration is dependant on environmental conditions. The other events: biomass production, reproduction, and death, depend only on internal parameters of a microbe (amount of biomass etc.) and $P_{D}$ which is not affected by environmental conditions. Therefore it is not necessary break chemical inflow / outflow further down to increment between each iteration step.

 In this process, we treat chemical levels as continuous but the microbes always treat the chemicals as units. So for a timestep, each iteration we might add 10.7 chemical units per iteration, but microbes in the system can only act on the integer amounts of chemicals present.

\subsection{Method}

We perform the following steps for each connectivity in list C:
\begin{enumerate}
\item Set up the planet's geological network
\begin{enumerate}
\item Begin the geological processes on the planet, allowing chemicals to build up
\item Seed planet with a single species when $T_{planet} = T_{pref}$ 
\item if $T_{pref}$ is never reached, seed after $5 \times 10^{4}$
\item The experiment ends after $5 \times 10^{5}$ timesteps after seeding
\end{enumerate}
\item Repeat  step b) 100 times with different random  seeds initialising the microbes
\item Repeat  steps  (a)  to (c)  100 times  with  different  random  seeds  initialising  the planet's geological network
\end{enumerate}

\subsection{Parameters}

Tables \ref{table:A2}, \ref{table:A3}, and \ref{table:A4} show the parameter values used to generate the data presented in this paper.

\begin{table}
\centering
\caption{Planet parameters}\label{table:A2}
 \begin{tabular}{ | p{1.3cm} | p{1.5cm} | p{4cm} | }
    \hline
    Parameter & Value & Description \\ \hline
    $N$ & 8 & Number of chemical species \\
    $N_{S}$ & 2 & Number of source chemistry \\
    $I_{N}$ & 75 & Rate  of chemical influx (units / timestep) per source chemical \\ 
    $O_{N}$ & 0.0001 & Rate  of chemical outflux (percentage / timestep) \\
    $P_{abiotic}$ & 1 & Probability of a chemical species having an insulating or reflecting effect  \\
    $a_{i}$ & [-1, +1] & A chemical species's reflective (if -ve) or insulating (if +ve) effect on the planet generated from range [-1, +1] \\
    $\beta_{star}$ & [500, 1000, 1500] & Solar radiation provided by host star / timestep \\
    $D$ & 75,000 & A constant to dampen the effects of a single `particle' in the atmosphere \\
    $C$ & [0.1, 0.2, 0.3, 0.4, 0.5, 0.6, 0.7, 0.8, 0.9] & Planet connectivity, i.e. the proportion of chemical species connected by geochemical processes  \\ \hline
     \end{tabular}
\end{table}

\begin{table}
\centering
\caption{Microbe parameters}\label{table:A3}
 \begin{tabular}{ | p{1.3cm} | p{1.5cm} | p{4cm} | }
    \hline
    Parameter & Value & Description \\ \hline
    $B_{R}$ & 120 & Reproduction threshold (biomass  units) \\
    $B_{D}$ & 50 & Starvation threshold (biomass  units) \\
    $P_{mut}$ & 0.01 & Probability of mutation at each locus during  reproduction \\ 
    $P_{D}$ & 0.002 &  Probability of death  by natural causes (other  than starvation) at each timestep \\
    $K^{max}$ & 10 & The maximum number of chemicals a microbe can eat per timestep when conditions are ideal  \\
    $N_{gene}$ & 8 & Microbe genome length \\
    $\lambda$ & 1 & Maintenance cost (biomass  units / timestep)  \\
    $\theta$ & 0.6 & Chemical  conversion efficiency  \\
    $\tau$ & 0.015 & Level of influence of abiotic  environment on metabolism  \\
    $T_{pref}$ & 1000 & Microbes' temperature preference  \\ \hline
     \end{tabular}
\end{table}

\begin{table}
\centering
\caption{Setup parameters}\label{table:A4}
 \begin{tabular}{ | p{1.3cm} | p{1.5cm} | p{4cm} | }
    \hline
    Parameter & Value & Description \\ \hline
    $M_{N}$ & 100 & Number  of individuals  in planet inoculum \\
    $B_{init}$ & 80 & Biomass of each seed individual \\
    $t_{run}$ & $5 \times 10^{5}$ & Duration of run (timesteps) \\ 
    $S_{W}$ & 0.002 &  Probability of death  by natural causes (other  than starvation) at each timestep \\
    $S_{F}$ & 1000 & Available food required for a seed species to be viable (units) \\ \hline
     \end{tabular}
\end{table}

\pagebreak
\section{Supplementary material}
\label{appendixB}

Here we present some results further exploring the ExoGaia model. These results do not change the main results of the paper but reinforce that ExoGaia exhibits self-regulation of planetary atmospheres by a microbial biosphere for a range of initial conditions.

\subsection{Reseeding with life}

\begin{figure}[htbp!]
\centering
    \begin{subfigure}{0.8\columnwidth}
        \centering
        \makebox[1.0\columnwidth][c]{\includegraphics[width=1.0\textwidth]{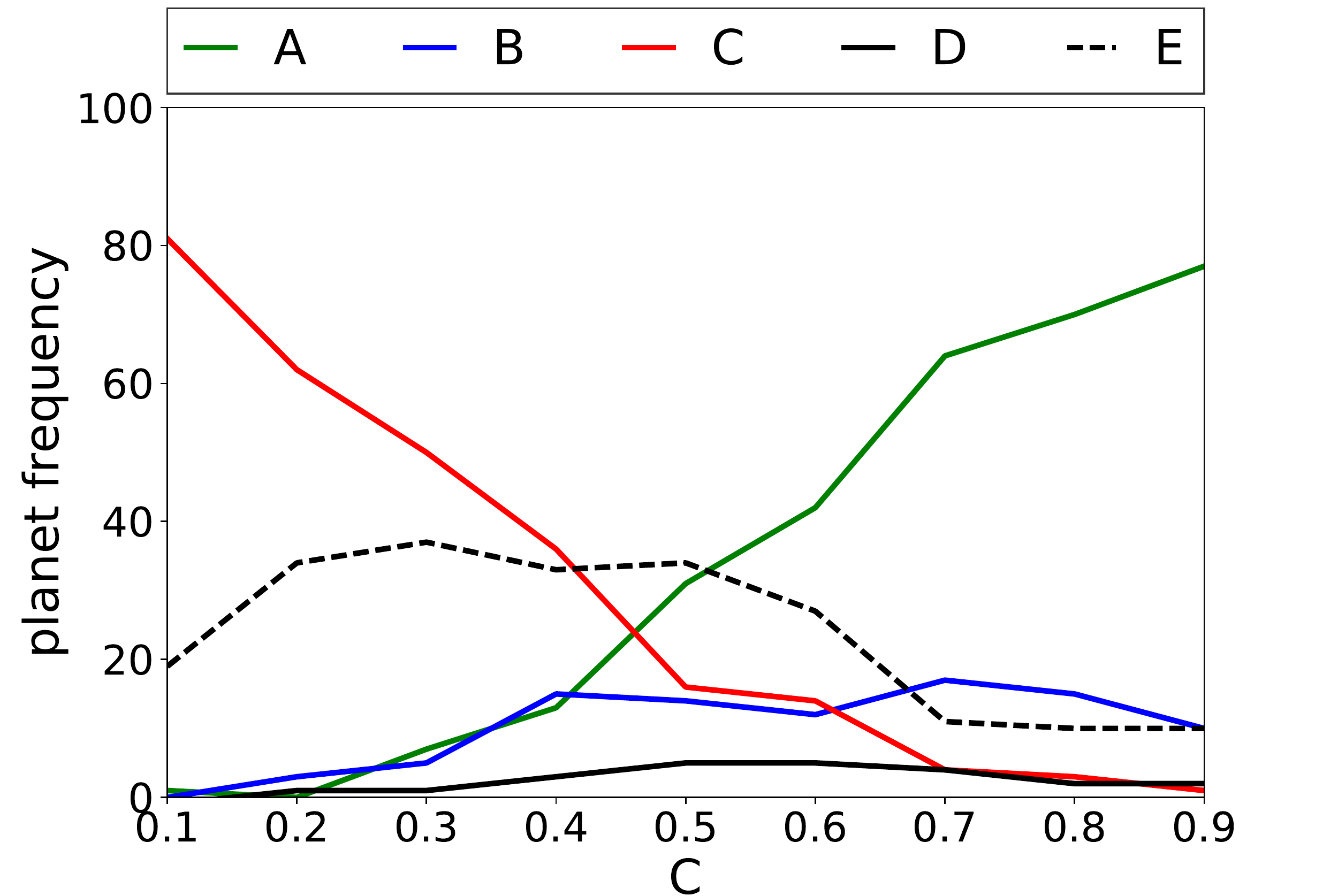}}
         \caption{No reseeding} \label{fig:B1a}
        \hspace*{5mm}
    \end{subfigure} 
        \begin{subfigure}{0.8\columnwidth}
        \centering
        \makebox[1.0\columnwidth][c]{\includegraphics[width=1.0\textwidth]{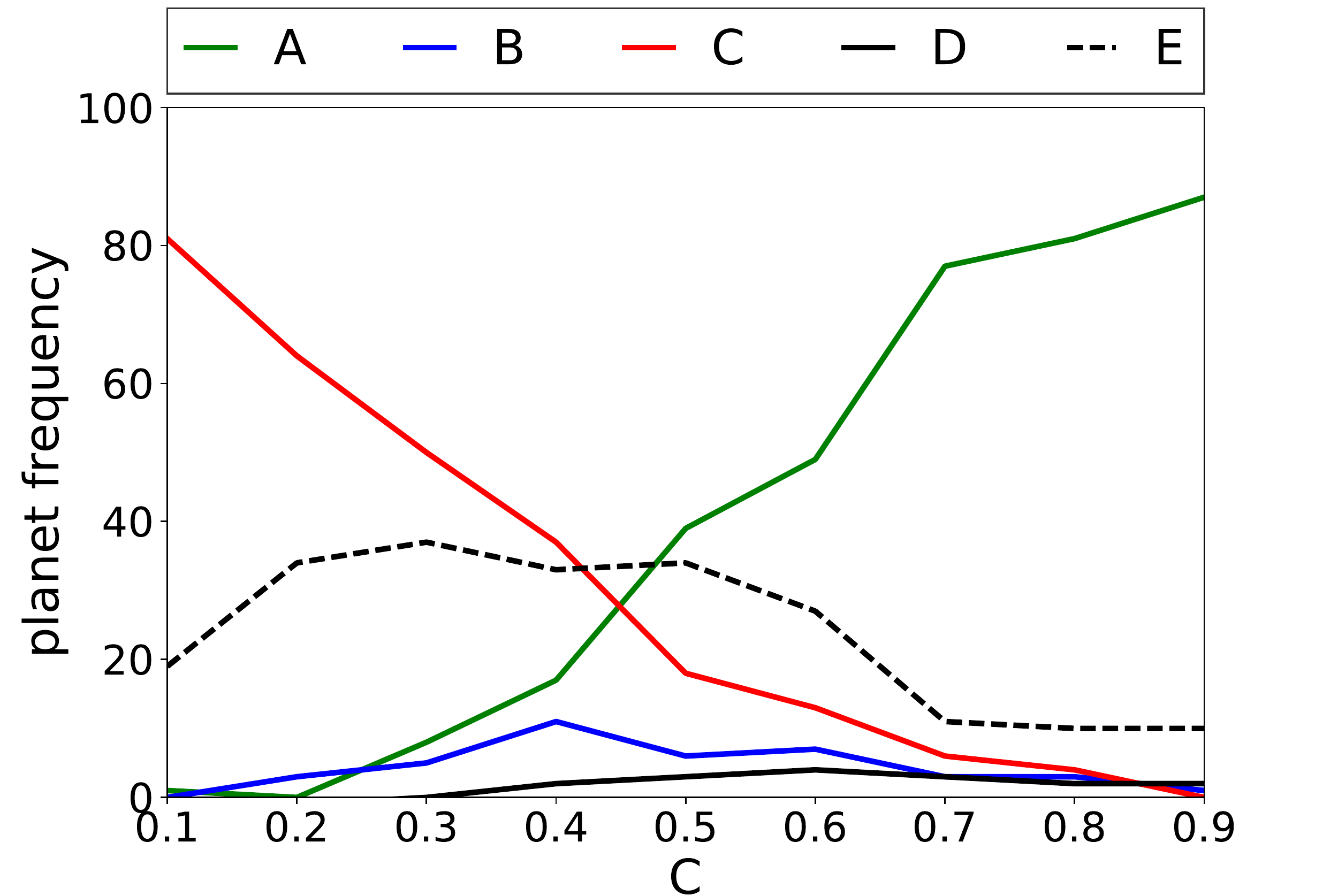}}
         \caption{With reseeding}
        \hspace*{5mm}
    \end{subfigure} 
\caption{The frequency of \textbf{A}biding, \textbf{B}ottleneck, \textbf{C}ritical, \textbf{D}oomed, and \textbf{E}xtreme planets against connectivity for Chemistry A.}
\label{fig:B1}
\end{figure}

We investigated the difference that reseeding with life had on the results. For the results presented thus far, once a system goes extinct, it remains so. We performed reseeding experiments where, after extinction, if the planet's temperature reached $T_{pref}$ again, the planet was reseeded with a single microbe species. We did not limit the number of times a system could be reseeded. As the origins of life remain largely a mystery we don't know whether life emerged once and took off straight away, or whether it required a few starts, so investigating each scenario is of interest. Another way to think about this reseeding is that some microbial `spores' may be so robust that they survive the crash of the system for `geological' lengths of time \cite{Nicholson2000} \cite{Wells2003} \cite{Wilkinson2006}, however with numbers so low that they do not have any influence on the evolution of their planet. Therefore if conditions improve, life is ready to take advantage immediately. For example it is speculated that, assuming Mars once had abundant life back when it had large quantities of liquid water, life could still exist on Mars in sparse pockets `waiting' for conditions to improve \cite{Wilkinson2006}. We found that the qualitative results are largely the same as for non-reseeding experiments, however survivability for all planets is improved. See Figure \ref{fig:B1} for a comparison between non-reseeding and reseeding experiments.

Figure \ref{fig:B1} shows the number of each Class of planets for Chemistry A both with and without reseeding. We see the most difference in the number of Bottleneck planets. For higher connectivity, we see the number of Bottleneck planets is lower for the reseeding experiment. These planets having now multiple chances for life to take hold have multiple attempts to overcome the bottleneck. This means some planets where bottlenecks were previously seen now become Abiding planets with all simulations surviving the experiment. 

Some previously Doomed planets became Critical planet class under the reseeding experiment, however most remained in the Doomed classification and those that transitioned to being critical planets were still poor long-term hosts for life, with the planet experiencing multiple reseeding events over the course of the experiment. Therefore, while some Doomed planets might, under reseeding, support life for overall longer timespans, if we consider the implications for real planets, we can infer that these planets would be unlikely to support complex life due to the frequent extinction events occurring.

\subsection{Chemical Set B and C}
\label{SectionB2}

We repeated  the non-reseeding experiments with two different chemical sets, to check that the results presented in the main body of the paper were not just a special case. We found that different chemical sets affect the quantitative results, but not the qualitative results of higher $C$ correlating with higher survival rates, and Gaian bottlenecks as an emergent feature of the model. The five planet classes emerge for all three chemical sets, with the number of Abiding planets increasing with increasing $C$, and the number of Critical planets decreasing with increasing $C$. The exact number of each planet class differs between chemical sets, however the key result is that chemical set A, used for the results in the main body of the paper, is not a special case.

\begin{table}
\centering
\caption{The greenhouse and albedo properties for chemical sets B and Chemistry C. The bold chemicals represent the influx chemical.}\label{table:B1}
 \begin{tabular}{ccc}
    \hline
    Chemical index & Chemical set B & chemical set C \\ \hline
    1 & -0.56 & -0.73 \\ 
    2 & 0.67 & 0.88 \\ 
    3 & 0.79 &  -0.82 \\ 
    4 & -0.40 &  \textbf{0.27} \\ 
    5 & \textbf{0.04} & 0.52 \\ 
    6 & \textbf{0.26} & 0.11 \\ 
    7 & 0.04 & \textbf{-0.38} \\ 
    8 & -0.31 & 0.39 \\ \hline
    Mean & 0.07 & 0.03 \\ \hline
    \end{tabular}
\end{table}

Table \ref{table:B1} shows the chemical species $a_{i}$ values for both chemical set B and C. We can see that we would expect a planet with chemical set B to be on average warmer than a planet with chemical set C, as the chemical species are on average more insulating. Both chemical sets are significantly cooler than chemical set A. We put these two different chemical sets though the same experiments as before, investigating only the non-reseeding case. We would predict that chemical set C, being colder than chemical set B, would result in fewer habitable planets.

\begin{figure}[htbp!]
\centering
    \begin{subfigure}{0.8\columnwidth}
        \centering
        \makebox[1.0\columnwidth][c]{\includegraphics[width=1.0\textwidth]{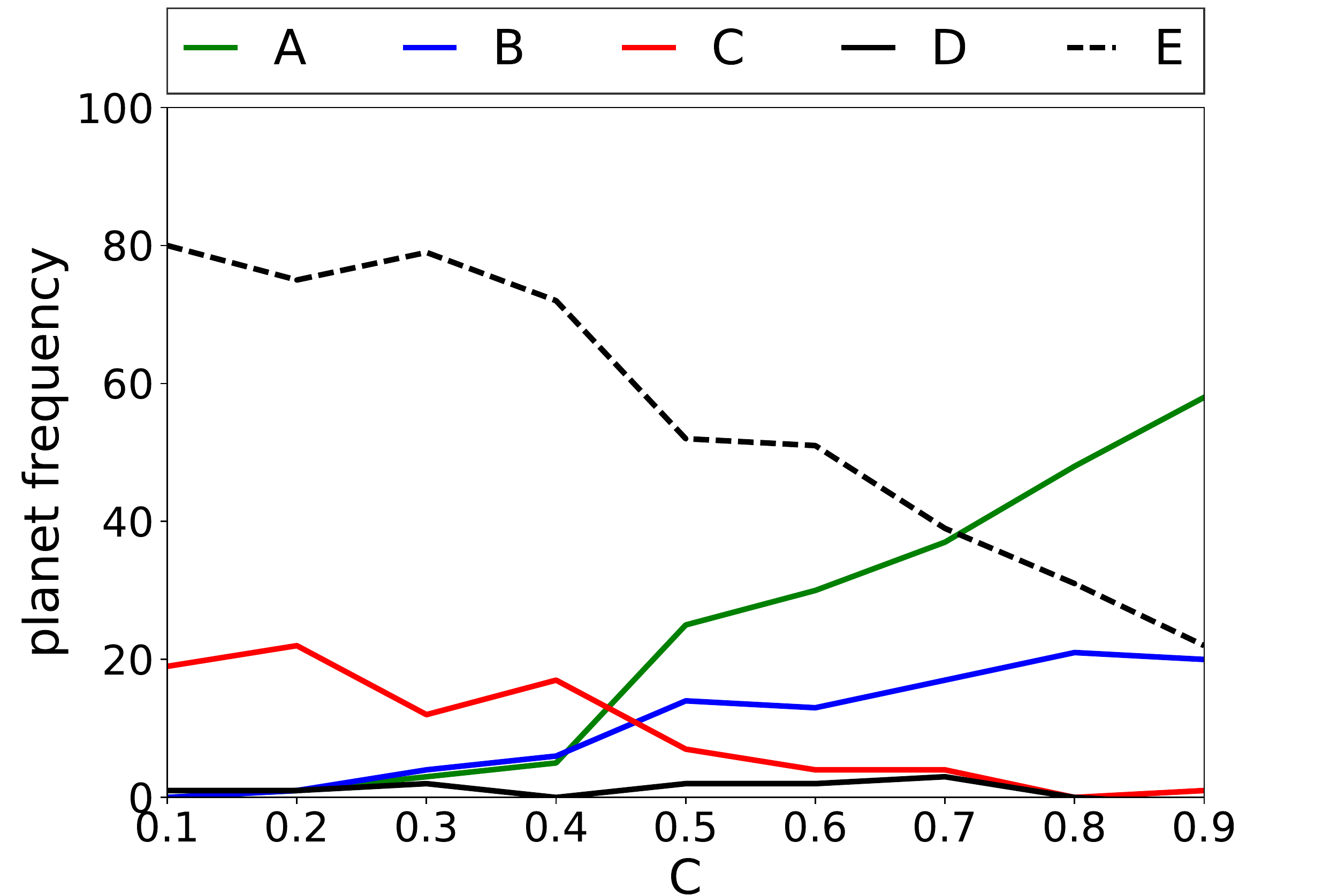}}
         \caption{Chemistry B}
        \hspace*{5mm}
    \end{subfigure} 
        \begin{subfigure}{0.8\columnwidth}
        \centering
        \makebox[1.0\columnwidth][c]{\includegraphics[width=1.0\textwidth]{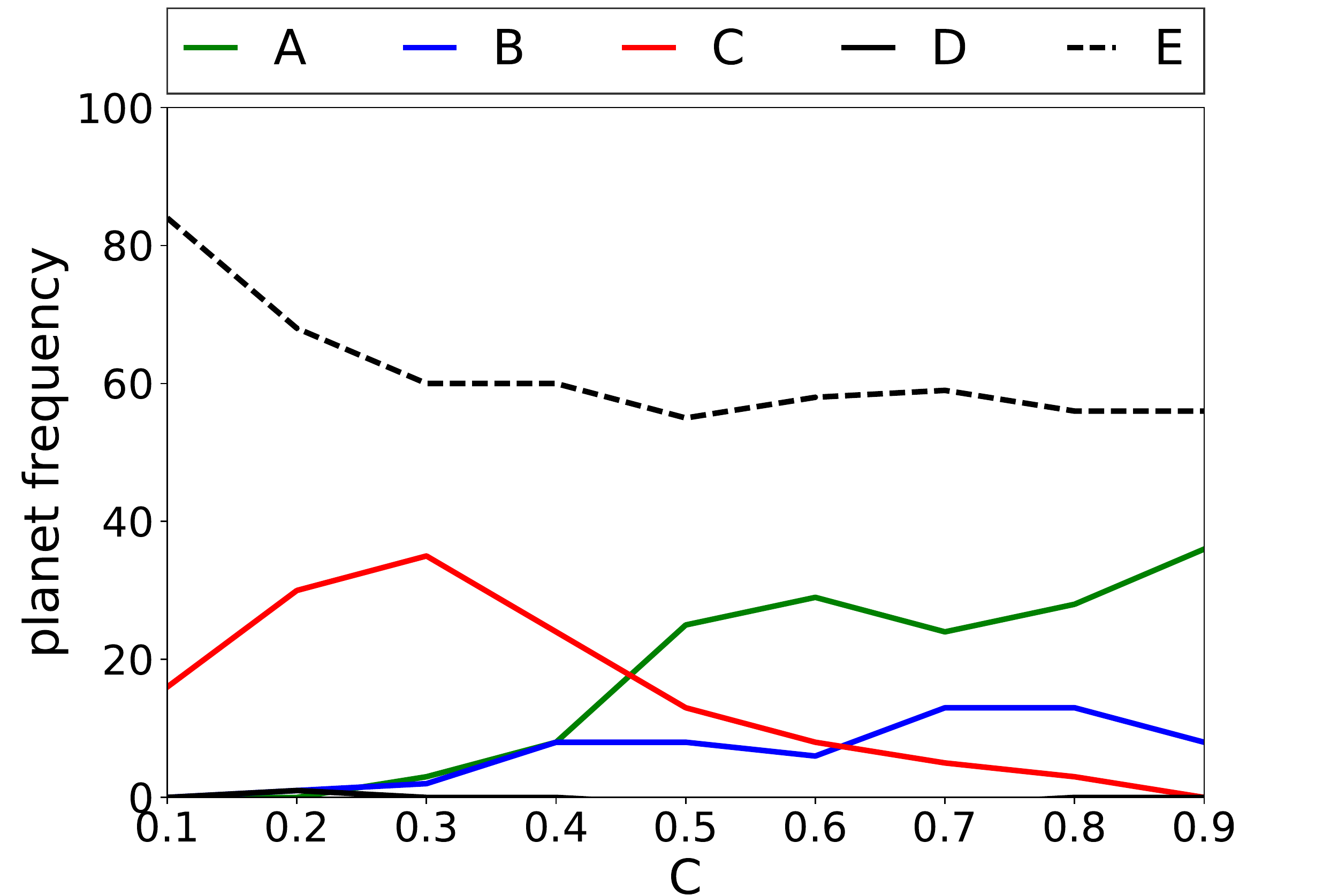}}
         \caption{Chemistry C}
        \hspace*{5mm}
    \end{subfigure} 
\caption{The frequency of \textbf{A}biding, \textbf{B}ottleneck, \textbf{C}ritical, \textbf{D}oomed, and \textbf{E}xtreme planets against connectivity}
\label{fig:B2}
\end{figure}

Figure \ref{fig:B2} shows the number of each class of planets for chemical set B and C. We see that these quantitatively differ from one another and from chemical set A, but general trends are similar. The number of Abiding planets is highest for high $C$. Critical and Extreme planets decrease with increasing $C$, and Bottleneck planets are more common for middling and high $C$ values. Doomed planets make up a very small percentage of the simulated planets for both chemical sets We find overall the colder chemical set C results in more Extreme planets, as expected. We see see that our original chemical set A, presented in the main body of the paper, is not a special case, and that viable biospheres are possible with different chemical sets. 

\subsection{Changing $\beta_{star}$}

The results presented so far have $\beta_{star} < \beta_{pref}$. This means that model planets need to have insulating atmospheres to reach habitable conditions. We now investigate how changing $\beta_{star}$ affects the results. We explore two cases: Chemical set D with $\beta_{star} = 1500$, therefore with $\beta_{star} > \beta_{pref}$ (instead of $\beta_{star} < \beta_{pref}$ as for chemical set A, B, and C), and chemical set E with $\beta_{star} = \beta_{pref} = 1000$. See Table \ref{table:B2} for the $a_{i}$ values for the chemical species of chemical sets D and E.

\begin{table}
\centering
\caption{The greenhouse and albedo properties for chemical sets D and Chemistry E. The bold chemicals represent the influx chemicals}\label{table:B2}
 \begin{tabular}{ccc}
    \hline
    Chemical index & Chemical set D & chemical set E \\ \hline
    1 & \textbf{0.95} & -0.88 \\ 
    2 & -0.70 & -0.91 \\ 
    3 & \textbf{-0.05} &  \textbf{-0.43} \\ 
    4 & -0.39 & 0.94 \\ 
    5 & -0.54 & 0.99 \\ 
    6 &  -0.20 & -0.04 \\ 
    7 & 0.78 & -0.80 \\ 
    8 & -0.19 & \textbf{-0.90} \\ \hline
    Mean & -0.34 & -2.03 \\ \hline
    \end{tabular}
\end{table}

\begin{figure}[htbp!]
\centering
    \begin{subfigure}{0.8\columnwidth}
        \centering
        \makebox[1.0\columnwidth][c]{\includegraphics[width=1.0\textwidth]{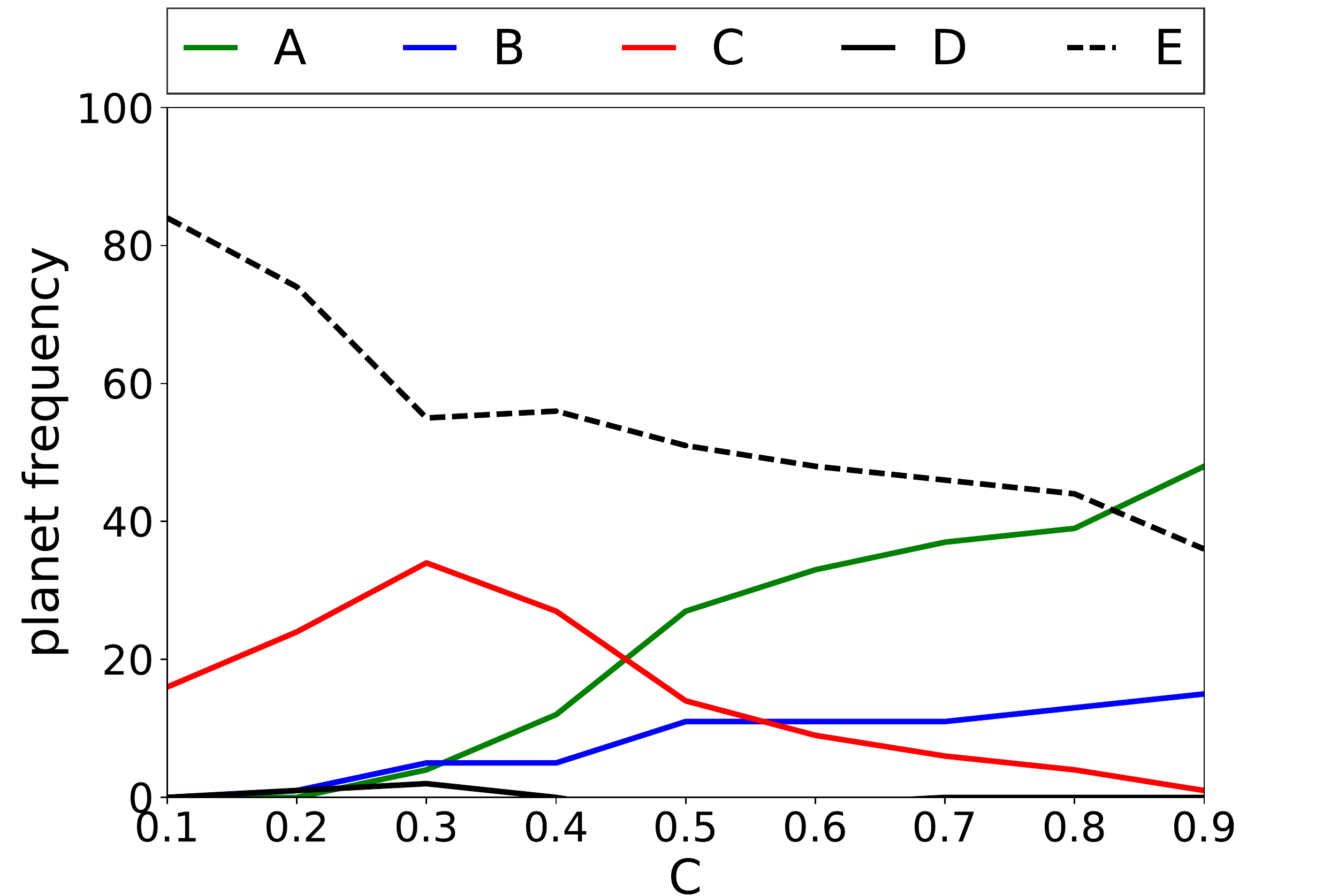}}
         \caption{Chemistry D with $\beta_{star} > \beta_{pref}$}\label{fig:B3a}
        \hspace*{5mm}
    \end{subfigure} 
        \begin{subfigure}{0.8\columnwidth}
        \centering
        \makebox[1.0\columnwidth][c]{\includegraphics[width=1.0\textwidth]{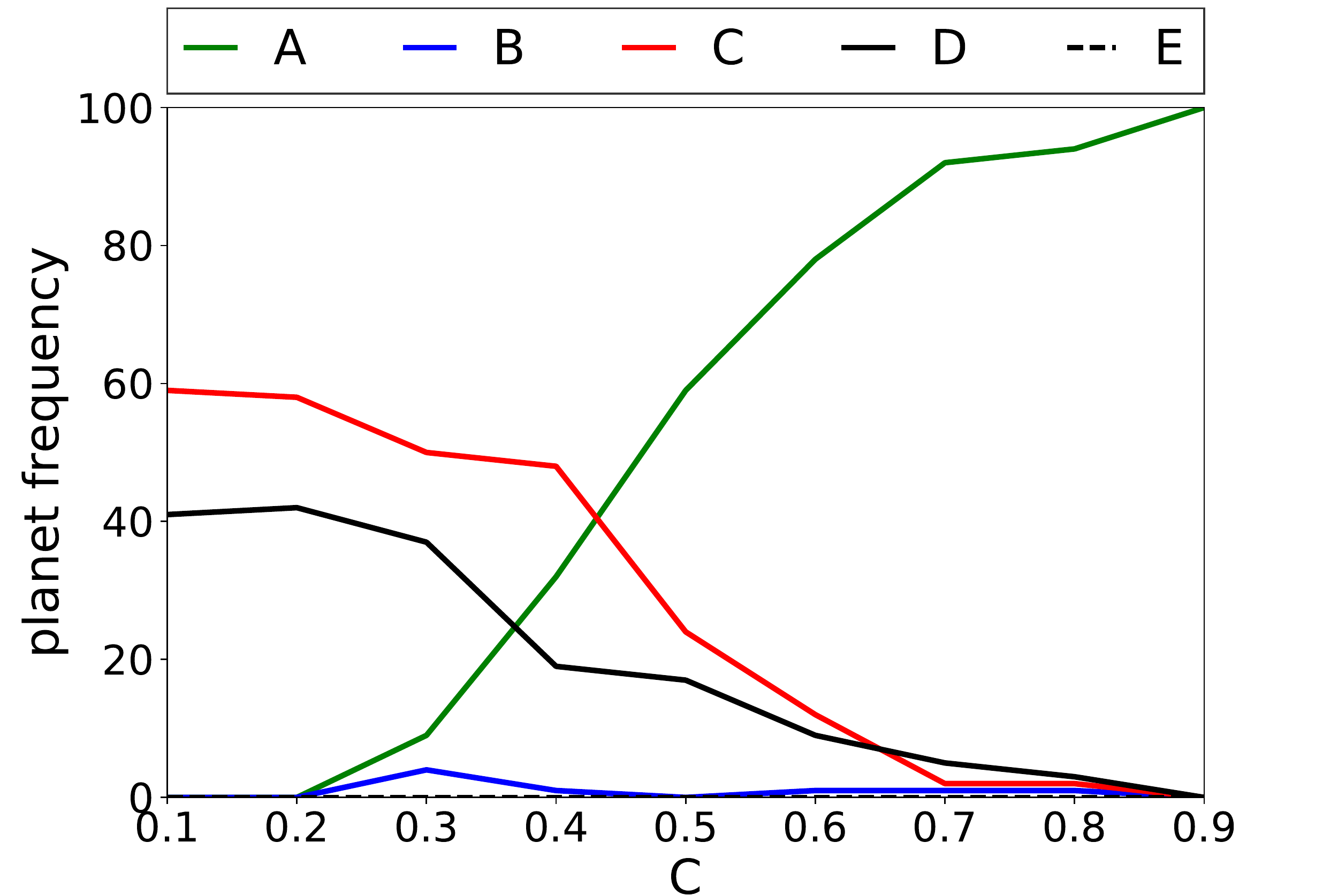}}
         \caption{Chemistry E with $\beta_{star} = \beta_{pref}$}\label{fig:B3b}
        \hspace*{5mm}
    \end{subfigure} 
\caption{The frequency of \textbf{A}biding, \textbf{B}ottleneck, \textbf{C}ritical, \textbf{D}oomed, and \textbf{E}xtreme planets against connectivity}
\label{fig:B3}
\end{figure}

We find for $\beta_{star} > \beta_{pref}$, 5 planet classes again emerge and temperature regulation can still take place. For a planet to be habitable when $\beta_{star} > \beta_{pref}$ the atmosphere must now have an overall cooling effect on the planet. In this scenario, rather than temperature regulation taking place below $T_{pref}$ with the microbes collectively reducing the insulating power of the atmosphere to maintain habitable conditions, regulation instead takes place above $T_{pref}$ with the microbes collectively reducing the reflective effect the atmosphere. The negative feedback loop, and the positive feedback loop are the same as outlined in Section 4.1 in the main paper but with the signs flipped such that 
when $T_{planet} < T_{pref}$, effects of increasing (+) the temperature are: 

\begin{enumerate}
\item + Temperature $\rightarrow$ + Population
\item + Population $\rightarrow$ + Temperature
\end{enumerate}

resulting in a runaway positive feedback loop. This also means decrease in $T_{planet}$ will result in a decrease in the microbe population, further decreasing $T_{planet}$ as abiotic processes dominate, leading to total extinction. A positive feedback loop where $T_{planet}$ is increasing will result in $T_{planet} > T_{pref}$ where the negative feedback loop occurs, as for $T_{planet} > T_{pref}$:

\begin{enumerate}
\item + Temperature $\rightarrow$ -Population
\item + Population $\rightarrow$ + Temperature
\end{enumerate}

resulting in a stabilising negative feedback loop. Figure \ref{fig:B3} shows the frequency of each planet class for chemical set D with $\beta_{star} > \beta_{pref}$ and for chemical set E with $\beta_{star} = \beta_{pref}$ . The $\beta_{star} > \beta_{pref}$ case qualitatively looks the same as the $\beta_{star} < \beta_{pref}$ scenarios for chemical set A, B, and C. All 5 planet classes are seen and as $C$ increases the long term habitability of planets increases.

Figure \ref{fig:B3b} shows planet class frequency for $\beta_{star} = \beta_{pref}$. The behaviour of the model changes slightly under these conditions. As a planet with no atmosphere will now have $T_{planet} = T_{pref}$, the microbes, once seeded, experience a positive feedback where an increase in population leads to an increase in habitability (as the atmospheric chemicals are reduced). Thus after seeding, if habitability prevails long enough, the microbes will quickly consume all of the atmosphere. With no atmosphere the population becomes nutrient limited and temperature regulation via a negative feedback loop does not take place. Microbes maintain habitable conditions simply by preventing any atmosphere building up. This phenomena is known as biotic-plunder \cite{Tyrrell2004} where the biota exhaust resources and so achieve stability, while the resources remain at very low levels. If microbe populations decrease due to stochastic fluctuations, atmospheric chemicals can build up, moving $T_{planet}$ away from $T_{pref}$ and leading to a positive feedback loop, resulting in extinction. For high $C$ the temperature change is small enough that microbe numbers can recover in time to consume the excess chemicals and remain nutrient limited. For low $C$, where chemical species can accumulate more rapidly, microbes are sometimes unable to prevent the positive feedback loop. Therefore we see far more Critical planets occurring for low $C$ than for high.

There are no Extreme planets when $\beta_{star} = \beta_{pref}$. As conditions at the start of each experiment have $T_{planet} = T_{pref}$, all planets spend time in a habitable temperature range and thus all planets are potentially habitable. Doomed planets are those where $T_{planet}$ diverges from $T_{pref}$ too quickly before a food source has built up for microbes, preventing successful colonisation of the planet. For higher $C$ where the chemicals are more evenly distributed between each chemical species, temperatures change at a slower rate, and thus the number of Doomed planets decreases.

Bottlenecks are rare for $\beta_{star} = \beta_{pref}$. Previous results, Figures 6 - 10 in the main paper, showed that when microbes were seeded on a planet when $\beta_{star} < \beta_{pref}$ they caused a reduction in habitability (the same is true for $\beta_{star} > \beta_{pref}$). For $\beta_{star} = \beta_{pref}$ the sudden decrease in atmospheric chemicals due to seeding will instead lead to an increase in habitability. This prevents much of the the bottleneck behaviour seen when $\beta_{star}$ is far from $\beta_{pref}$, as in Figures 7 and 8, i.e. the decrease in habitability followed by a rapid population reduction, with the population sometimes recovering and sometimes going extinct. Bottleneck behaviour can still emerge however when microbes do not evolve metabolisms fast enough to consume all chemical species building up in the atmosphere. However, with the increase in habitability after seeding, reproduction rates and thus mutation rates are higher than when $\beta_{star}$ is far from $\beta_{pref}$ and so varied metabolisms appear more rapidly making Bottleneck planets less likely for $\beta_{star} = \beta_{pref}$.

It seems unlikely that the biosphere on a real planet would consume the entire atmosphere, perhaps making the model results for $\beta_{star} = \beta_{pref}$ less realistic than for $\beta_{star} < \beta_{pref}$ or $\beta_{star} > \beta_{pref}$, however nutrient limitation is a well-known phenomena in ocean systems \cite{Moore2013}. The ExoGaia model demonstrates that habitable conditions can be maintained by a biosphere under a range of conditions, and we see that for each scenario tested the underlying geochemical network plays a key role in determining a planet's suitability for long term habitability.

\subsection{Changing the $\beta$ and $T$ relation}

\begin{figure}[htbp!]
\centering
    \begin{subfigure}{0.45\columnwidth}
        \centering
        \makebox[1.0\columnwidth][c]{\includegraphics[width=1.0\textwidth]{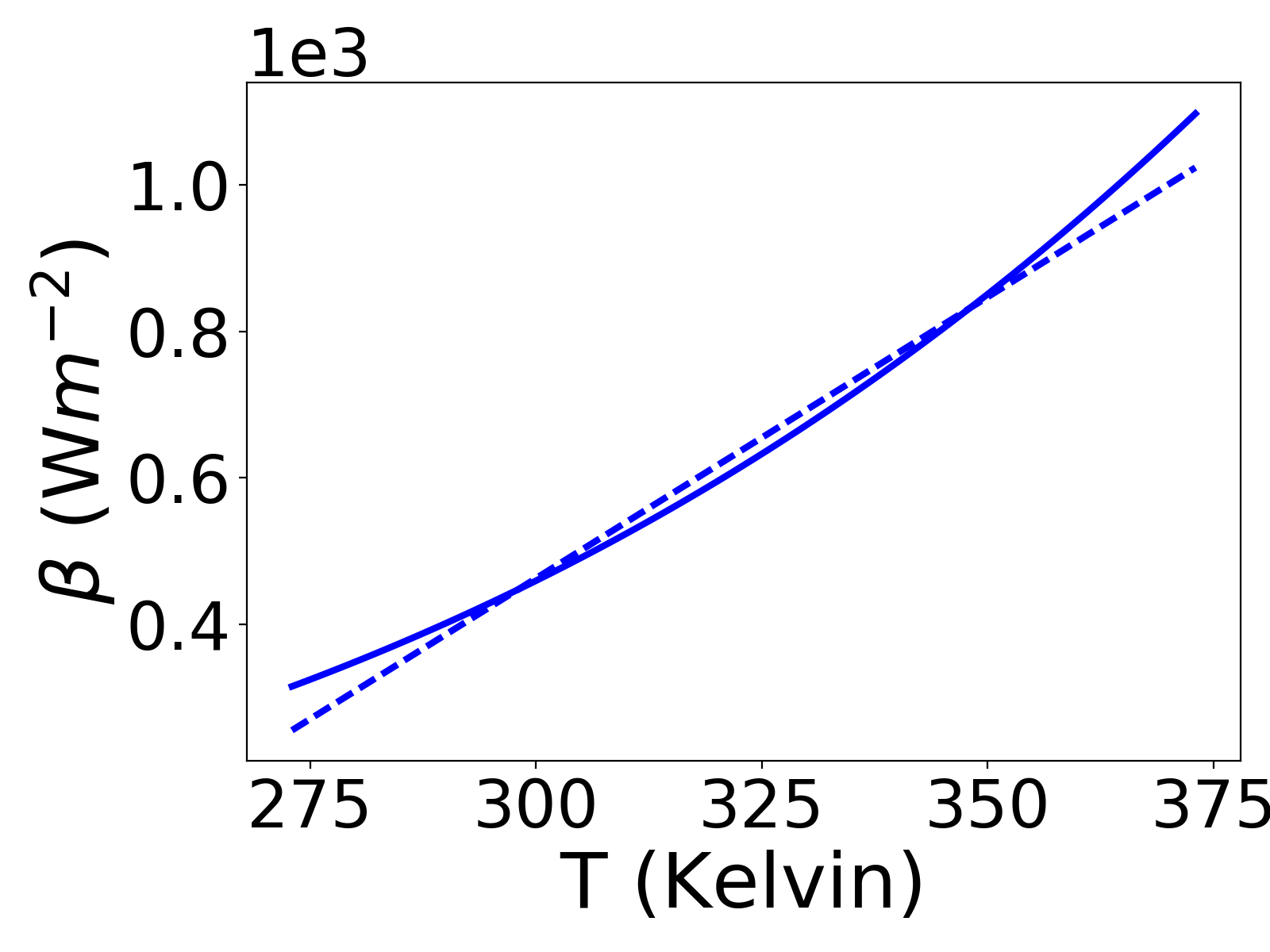}}
         \caption{Real world habitable temperature range}\label{fig:B4a}
        \hspace*{5mm}
    \end{subfigure} 
        \begin{subfigure}{0.45\columnwidth}
        \centering
        \makebox[1.0\columnwidth][c]{\includegraphics[width=1.0\textwidth]{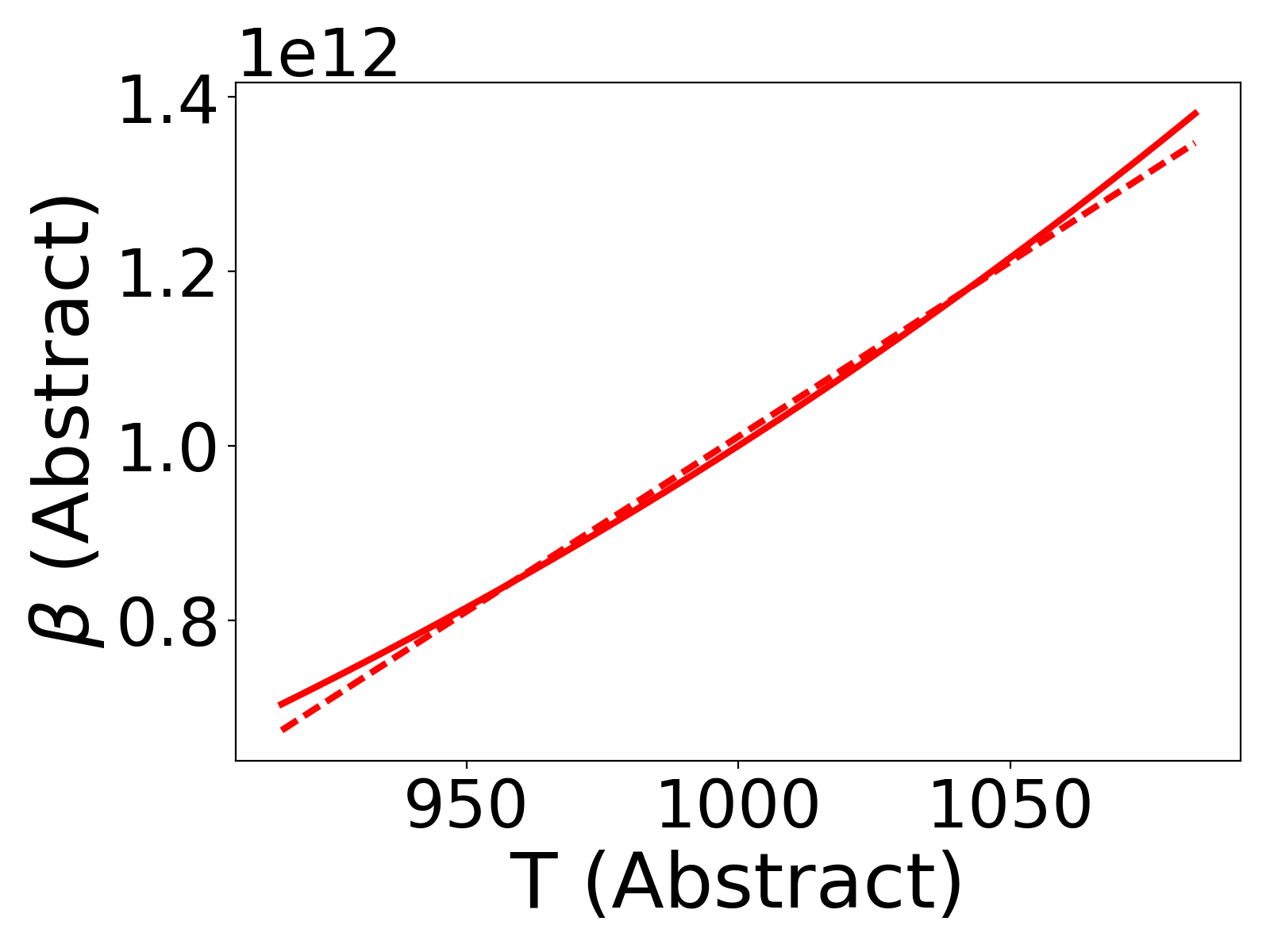}}
         \caption{ExoGaia habitable temperature range}\label{fig:B4b}
        \hspace*{5mm}
    \end{subfigure} 
\caption{$\beta \sim T^{4}$ (solid) and linear approximations (dashed) for the habitable temperature ranges for the real world (a) and ExoGaia worlds (b)}
\label{fig:B4}
\end{figure}

The results presented throughout the main paper and this Appendix thus far have used a linear relationship between $\beta$ and temperature. Figure \ref{fig:B4a} shows a plot of the Stefan-Boltzmann law: $\beta = \sigma T^{4}$ (where $\sigma$ is the Stefan-Boltzmann constant), for temperatures between $0 - 100 ^{o}C$, and a linear approximation (dashed). We can see that the linear approximation is a close fit to the $T^{4}$ curve. Figure \ref{fig:B4b} shows $\beta = T^{4}$ for the range of habitable abstract temperatures in ExoGaia: $916 - 1084$, also with a linear approximation (dashed). The $T^{4}$ relation in Figure \ref{fig:B4b} is slightly less curved than in Figure \ref{fig:B4a} however they are not vastly different. Therefore a linear approximation of $\beta \sim T$ in these temperature ranges is a close approximation.

We can investigate the behaviour of the ExoGaia model with the more realistic $\beta = T^{4}$ instead of $\beta = T$. Multiplying $T^{4}$ by a constant, e.g. $\sigma$, is not important as this constant cancels out in the equation updating the thermal energy in a planet's atmosphere (Equation \ref{equation:A13}) and so can be safely ignored. Omitting $\sigma$ also serves as a reminder that all temperatures in ExoGaia are abstract.

\begin{figure}[htbp!]
\centering
    \begin{subfigure}{0.8\columnwidth}
        \centering
        \makebox[1.0\columnwidth][c]{\includegraphics[width=1.0\textwidth]{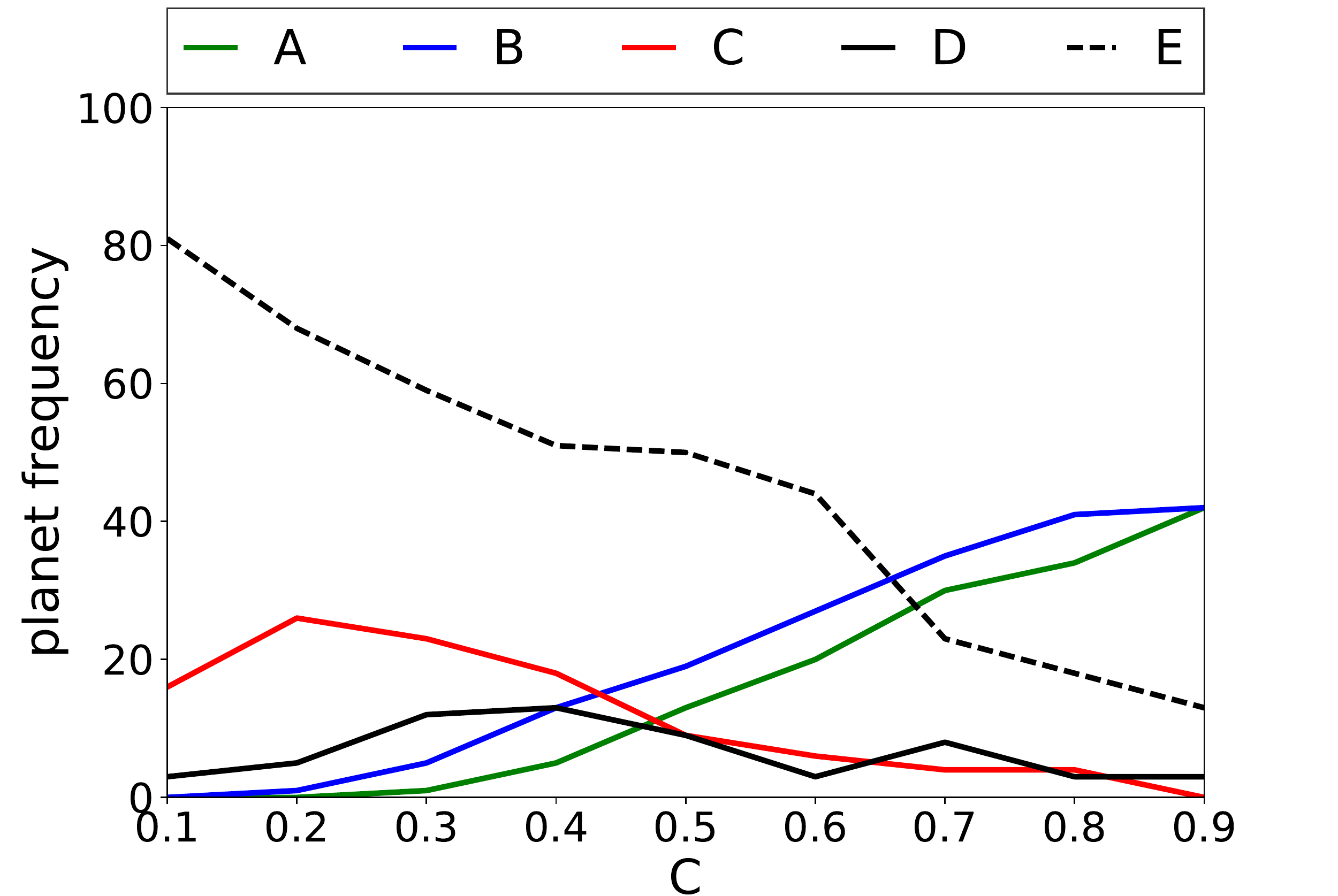}}
        \hspace*{5mm}
    \end{subfigure} 
\caption{The frequency of \textbf{A}biding, \textbf{B}ottleneck, \textbf{C}ritical, \textbf{D}oomed, and \textbf{E}xtreme planets against connectivity for simulations where $\beta = T^{4}$.}
\label{fig:B5}
\end{figure}

Figure \ref{fig:B5} shows the frequency of each planet class against connectivity. These results use the same parameters as those used for results in the main paper, the only change being $\beta = T \rightarrow \beta = T^{4}$. We see that the overall behaviour of the model is unchanged, 5 planet classes emerge, with increasing connectivity correlated with increased long-term habitability success for planets. The planet class frequencies between Figure \ref{fig:B5} and \ref{fig:B1a} (the results from the main body of the paper) differ significantly however.

The reason for this is not due to the curved $\beta$ and $T$ relation, but due to the fact that when $\beta = T^{4}$, doubling $\beta$ no longer corresponds to doubling $T$. The $\beta = T^{4}$ relation results in far more planets being too cold for life, hence the large number of Extreme planets seen in Figure \ref{fig:B5}. As temperature in the ExoGaia model is unconstrained, fitting $\beta = A \times T^{4} + B$ to approximate $\beta = T$ (to capture the curvature of a $T^{4}$ relation but maintain similarity with the original data), results in imaginary temperatures being possible - which is of course unphysical. Therefore, no fitting was performed and thus the relative planet class frequencies are quite different. However Figure \ref{fig:B5} demonstrates that using a $T^{4}$ relation instead of a linear one does not impact the important results of the model, and that the model results are robust to significant changes to the $\beta$ and $T$ relationship.




\label{lastpage}
\end{document}